\documentclass[prd,preprint,showpacs,amssymb,superscriptaddress,nofootinbib]{revtex4}

\usepackage{color}
\usepackage{graphicx}

\newcommand{\ba}{\begin{eqnarray}}
\newcommand{\ea}{\end{eqnarray}}

\newcommand{\be}{\begin{equation}}
\newcommand{\ee}{\end{equation}}

\begin{document}

\preprint{SLAC-PUB-10990, WM-05-102, JLAB-PHY-05-297}  
\vglue 5mm

\title{
The two-photon exchange contribution to
elastic electron-nucleon scattering at large momentum transfer}
%
\author{Andrei V. Afanasev}
\email[]{afanas@jlab.org}
\affiliation{Thomas Jefferson National Accelerator Facility, 
Newport News, VA 23606, USA}

\author{Stanley J. Brodsky}
\email[]{sjbth@slac.stanford.edu}
\affiliation{SLAC, Stanford University, Stanford, CA 94309, USA}

\author{Carl~E.~Carlson}
\email[]{carlson@physics.wm.edu}
\affiliation{Department of Physics, College of William and Mary,
Williamsburg, VA 23187, USA}
\email[]{carlson@physics.wm.edu}

\author{Yu-Chun Chen}
\email[]{snyang1@phys.ntu.edu.tw}
\affiliation{Department of Physics, National Taiwan University, 
Taipei 10617, Taiwan}

\author{Marc Vanderhaeghen}
\email[]{marcvdh@jlab.org}
\affiliation{Thomas Jefferson National Accelerator Facility, 
Newport News, VA 23606, USA}
\affiliation{Department of Physics, College of William and Mary,
Williamsburg, VA 23187, USA}
\date{January 31, 2005}
\begin{abstract}
We estimate the two-photon exchange contribution to elastic
electron-proton scattering at large momentum transfer by using a
quark-parton representation of virtual Compton scattering.  We thus can
relate the two-photon exchange amplitude to the generalized parton
distributions which also enter in other wide angle scattering
processes. We find that the interference of one- and two-photon
exchange contribution is able to substantially resolve the difference
between electric form factor measurements from Rosenbluth and
polarization transfer experiments.   Two-photon exchange has additional
consequences which could be experimentally observed, including nonzero
polarization effects and a positron-proton/electron-proton scattering
asymmetry.   The predicted Rosenbluth plot is no longer precisely
linear; it acquires a measurable curvature, particularly at large
laboratory angle.   
\end{abstract}
\maketitle


\section{Introduction}


There are two experimental methods for extracting the ratio of
electric ($G_E^p$) to magnetic ($G_M^p$) proton form factors from electron-proton scattering:
unpolarized measurements employing the Rosenbluth separation technique,
and polarization experiments. In the latter case, one measures the correlation of the spin of  the incident polarized electron with the polarization components of the outgoing proton, parallel $P_l$
or perpendicular $P_s$ (in the scattering plane) to its momentum~\cite{Jones00,Gayou02,Arn81}.  The ratio of cross sections for the two outgoing proton polarizations gives  $G_E/G_M$ directly:
\ba
{P_s \over P_l} = 
		- \sqrt{2\varepsilon \over \tau (1+\varepsilon)}\  
		{ G_E(Q^2)  \over G_M(Q^2)
		}  \ .
\ea
The kinematic functions $\varepsilon$ and $\tau$ are
\ba
 \tau \equiv {Q^2 \over 4M^2}   \ , 
\ea
									and
\ba
{1 \over \varepsilon} \equiv 1 + 2(1+\tau) \tan^2 {\theta\over 2} \ ,
\ea
where $Q^2=-q^2=-t$ is the momentum transfer squared, $\theta$ is the laboratory scattering angle, and $0 \le \varepsilon \le 1$. Equivalent information may be obtained in scattering of longitudinally polarized
electrons on a polarized proton target.

The Rosenbluth method relies on measuring the differential cross section\ba
{d\sigma \over d\Omega_{Lab}} \propto
		 G_M^2 + \frac{\varepsilon}{\tau} G_E^2 
				\ ,
\ea
with the proportionality factor being well known, and isolating the $\varepsilon$ dependent term. 
In each case, the extraction method for $G_E/G_M$ assumes single-photon exchange between the electron and nucleon.

Recent polarization experiments at  the Thomas Jefferson Laboratory (JLab)~\cite{Chr04,Arr03} have confirmed the
earlier Rosenbluth measurements from SLAC ~\cite{Slac94}.  However,  at large $Q^2$, all of  the Rosenbluth measurements are at distinct variance with JLab measurements of $G^p_E/ G^p_M$ obtained using the polarization technique~\cite{Jones00,Gayou02}.   Since $G_E^p$
contributes to the unpolarized cross section at only a few percent level for
the $Q^2$ range in question, it is necessary to identify any possible systematic corrections to the Rosenbluth measurements at the percent level which could be responsible for this discrepancy.

One possible explanation for the discrepancy between the Rosenbluth and polarization methods is the presence of  two-photon
exchange effects, beyond those which have already been accounted for in the standard treatment of
radiative corrections. A general
study of two- (and multi)-photon exchange contributions to the
elastic electron-proton scattering observables was given in~\cite{GV03}.  In that work, it was noted that the interference of the two-photon exchange
amplitude with the one-photon exchange amplitude could
be comparable in size to the $(G^p_E)^2$ term in the unpolarized cross
section at large $Q^2$.  In contrast, the two-photon exchange effects do not impact the
polarization-transfer extraction of $G_E/G_M$ in an equally significant way.  Thus a missing and
unfactorizable part of the two-photon exchange amplitude at the level
of a few percent may well explain the discrepancy between the two
methods.

Realistic calculations of
elastic electron-nucleon scattering beyond the Born approximation are
required in order to demonstrate in a quantitative way that two-photon exchange effects are
indeed able to resolve this discrepancy.   In particular, one wants to study quantitatively the ``hard'' corrections  which will arise when both exchanged photons are far off shell or the intermediate nucleon state suffers inelastic excitations.   Calculations of  these corrections require a knowledge of the internal structure of the nucleon and thus could not be included in the classic~\cite{oldyennie,MoTsai68} and were not included in the more recent~\cite{MT00,Andrei01} calculations of radiative corrections to $eN$ elastic scattering.

A first step was performed recently in~\cite{BMT03}, where the
contribution to the two-photon exchange amplitude was calculated for the elastic
nucleon intermediate state. In that calculation it was found that the two-photon
exchange correction with an intermediate nucleon has the proper sign and
magnitude to partially resolve the discrepancy between the two
experimental techniques. 

In an earlier short note~\cite{YCC04}, we reported the first
calculation of the hard two-photon elastic electron-nucleon scattering amplitude at large
momentum transfers by relating  the required virtual Compton process on the nucleon to
generalized parton distributions (GPD's) which also enter in other wide
angle scattering processes.
This approach effectively sums all possible excitations of inelastic
nucleon intermediate states.  We  found that the two-photon corrections
to the Rosenbluth process indeed can substantially reconcile the two
ways of measuring $G_E/G_M$.  Our goal in this paper is to give a
detailed account of our work, and to present numerical results for a
number of quantities not included in the shorter report.

Perturbative QCD  factorization methods for hard exclusive processes provide a systematic method for computing the scaling and angular dependence of real and virtual Compton scattering at large $t$.
For example, PQCD
predicts that the leading-twist amplitude for Compton
scattering $\gamma p \to \gamma p$  can be factorized as a product of
hard-scattering amplitudes $T_H(\gamma qqq \to \gamma qqq),$ where the
quarks in each proton are collinear, convoluted with the initial and
final proton distribution amplitudes
$\phi(x_i,Q)$~\cite{Brodsky:1981rp}.  All of the hard-scattering
diagrams fall at the same rate at large momentum transfer whether or
not the photons interact on the same line.  Although, the predictions for the power-law falloff and angular dependence of Compton scattering are consistent with experiment,  the  leading-twist PQCD calculations of the wide angle Compton amplitudes appear to substantially underpredict the magnitude of the observed Compton cross sections~\cite{Brooks:2000nb}.

Since an exact QCD analysis of virtual Compton scattering does not appear practical,
we have modeled the hard two-photon exchange amplitude using the
``handbag approximation"~\cite{Brodsky:1973hm}, in which both photons
interact with the same quark. The struck quark is treated as
quasi-on shell. In particular, we have neglected the amplitudes where
the two hard protons connect to different quarks, the ``cat's ears"
diagrams, as well as the diagrams in which gluons interact on the
fermion line between the two currents.   The handbag diagrams contain the 
``$J=0$" fixed pole, the essential energy-independent contribution to the real part of 
Compton amplitude which arises due to the local structure of the quark current~\cite{Damashek:1969xj,Brodsky:1971zh}. The
handbag approximation has proven phenomenologically successful in
describing wide-angle Compton scattering at moderate energies and
momentum transfers.   As we shall show, the handbag approximation 
allows the two-photon exchange amplitude to be linked to the generalized
parton distributions (GPD's)~\cite{Die99,Rad98,Huang:2001ej}, thus providing considerable phenomenological guidance.

Brooks and Dixon
and Vanderhaeghen {\it et al.}~\cite{Brooks:2000nb} have shown that PQCD
diagrams where the photons attach to the same quark dominate the
Compton amplitude on the proton, except at backward
center-of-mass angles.\footnote{ This suppression of the cat's ears diagrams at forward angles 
could be due to the momentum mismatches which occur when photons couple
to different quarks. Another
possible explanation is that in some kinematic regions,
the cat's ears and handbag amplitudes have the same magnitude (or nearly so) 
except for the charge factors.  In these regions, the Compton amplitude
would be proportional to the total charge squared $(2 e_u + e_d)^2$ of
the target proton.     This is precisely the case in the low energy
limit, where the Compton amplitude is indeed proportional to $(2 e_u +
e_d)^2= 1$ for a proton.  The result  is reproduced by the handbag
diagrams alone since, coincidentally, $2 e^2_u + e_d^2 = 1$.   In this
scenario, the handbag approximation will fail for Compton scattering on
a neutron or deuteron target.  At higher energies, discussion of this
scenario pertains to large angle Compton scattering, since in the
forward direction the handbag diagrams are known to dominate.}.
The dominance of the 
handbag diagrams in the PQCD analysis provides some justification for the 
the use of the handbag approximation.  However, it should be noted that 
Gunion and Blankenbecler~\cite{Gunion:1970yy} have shown that
electron-deuteron scattering is  dominated by the cat's ears
diagrams at large momentum transfer provided that the deuteron wave function has Gaussian fall-off.   The dominance of the handbag diagrams thus  depends on the nature of  the QCD wave functions, and the precise situation in the present case remains a subject for future 
study.

Recently, a new category of Rosenbluth data has become available where the recoiling proton is detected~\cite{Qattan:2004ht}.  The new data appear to confirm the older data, where the scattered electron was detected. The two-photon exchange contributions are the same whatever particle is detected.  However, the bremsstrahlung corrections, which are added to obtain an infrared finite result, are different.  We shall defer detailed discussion of the proton-detected data until we can reevaluate the original proton-observed electron-proton bremsstrahlung interference calculations~\cite{oldyennie,oldkrass} as well as  examine the radiative corrections which have been applied to the new data~\cite{Ent:2001hm,Qattan:2004ht}. 

The plan of this paper is as follows:

The next section is devoted to kinematics, including the definitions of the  invariants which define the scattering amplitudes and the formulae for the cross sections and polarizations in terms of those invariants.  There are choices in the definitions of the invariants.  We have presented the bulk of the paper with one choice; a sometimes useful alternative choice is summarized with cross section and polarization formulas in Appendix~\ref{sec:axial}.  Section~\ref{sec:3} gives analytic results for the two-photon exchange scattering amplitudes at the  electron-quark level, the  hard scattering amplitudes required for the partonic calculation of two-photon exchange in electron-nucleon scattering.  We have generally treated the quarks as massless.  A quantitative discussion of modifications following from finite quark mass appears in Appendix~\ref{sec:withmass}.  Section~\ref{sec:4} details the embedding of the partonic amplitude within the nucleon scattering amplitude, using dominance of handbag amplitudes and GPD's.  This section also discusses the particular GPD's  which we have used in our numerical calculations.  Section~\ref{sec:5} shows numerical results, given graphically, for cross sections, single spin asymmetries, polarization transfers, and positron-proton vs. electron-proton comparisons.  Section~\ref{sec:5} also includes commentary about the possibility of extending the calculations to backward scattering (small values of $|u|$), and an assessment of how well  two-photon physics reconciles the Rosenbluth and polarization transfer measurements of $G_E/G_M$.  Section~\ref{sec:theend} summarizes our conclusions.


\section{Elastic electron-nucleon scattering observables}

\label{sec:observables}


In order to describe elastic electron-nucleon
scattering,
\begin{equation}
										\label{Eq:intro.2}
	l(k,h)+N(p,\lambda_N)\rightarrow l(k',h')+N(p',\lambda'_N),
\end{equation}

\noindent where $h$, $h'$, $\lambda_N$, and $\lambda'_N$ are helicities, we adopt the definitions
\begin{equation}
										\label{Eq:intro.3}
	P=\frac{p+p'}{2},\, K=\frac{k+k'}{2},\, q=k-k'=p'-p \ ,
\end{equation}

\noindent  define the Mandelstam variables
\ba
s = (p+k)^2,  \quad
t = q^2 = - Q^2,   \quad
u = (p-k')^2,
\ea

\noindent let $\nu \equiv K\cdot P$, and let $M$ be the nucleon mass.  

The $T$-matrix helicity amplitudes are given by
\ba
T^{h',h}_{\lambda'_N, \lambda_N} \equiv
	\left\langle k', h'; p', \lambda'_N \right| T 
		\left| k, h; p, \lambda_N \right\rangle  \ .
\ea
Parity invariance reduces the number of independent helicity amplitudes from 16 to 8.  Time reversal invariance further reduces the number to 6~\cite{Goldb57}.  Further still, in a gauge theory lepton helicity is conserved to all orders in perturbation theory when the lepton mass is zero.  We shall neglect the lepton mass.  This finally reduces the number of independent helicity amplitudes to 3, which one may for example choose as
\ba
T^{+,+}_{+,+} \ ; \quad T^{+,+}_{-,-} \ ; \quad 
	T^{+,+}_{-,+} = T^{+,+}_{+,-} \ .
\ea
(The phase in the last equality is for particle momenta in the $xz$ plane, and is valid whether we are in the center-of-mass frame, the Breit frame, or the symmetric frame to be defined below.)

Alternatively, one can expand in terms of a set of three independent Lorentz structures, multiplied by three generalized form factors.  Only vector or axial vector lepton currents can appear in order to ensure lepton helicity conservation.   A possible $T$-matrix expansion is (removing an overall energy-momentum conserving $\delta$-function),
\begin{eqnarray}						\label{eq:alt}
T_{h, \, \lambda'_N \lambda_N} = {e^2 \over Q^2 }  &\Bigg\{&
		\bar u(k',h) \gamma_\mu u(k,h) \times 
		\bar u(p',\lambda'_N) \left[ \gamma^\mu G'_M - {P^\mu \over M} 		
			F'_2 \right] u(p,\lambda_N)
							\nonumber \\
		&+& \bar u(k',h) \gamma_\mu \gamma_5 u(k,h) \times 
		\bar u(p,\lambda'_N) \, \gamma^\mu \gamma^5 G'_A \,  u(p,\lambda_N)
						\ \ 	  	\Bigg\}  \ .
\end{eqnarray}
This expansion is general.  The overall factors and the notations $G'_M$ and $F'_2$ have been chosen to have a straightforward connection to the standard form factors in the one-photon exchange limit.  

There is no lowest order axial vector vertex in QED: the effective axial vertex in the expansion arises from multiple photon exchanges and vanishes in the one-photon exchange limit.
One may eliminate the axial-like term using the identity
\ba														\label{eq:theorem}
\bar{u}(k') \gamma\cdot P u(k)  \times  \bar{u}(p') \gamma\cdot K u(p)
		&=&
		{s-u\over 4} \,
		\bar{u}(k') \gamma_\mu u(k)  \times  \bar{u}(p') \gamma^\mu u(p)
								\nonumber \\
		&+&  {t\over 4} \,
		\bar{u}(k') \gamma_\mu \gamma_5 u(k)  \times 
		\bar{u}(p') \gamma^\mu \gamma^5 u(p) \ ,
\ea
which is valid for massless leptons and any nucleon mass.  Hence, an equivalent $T$-matrix expansion is
\ba														\label{eq:tmatrix}
T_{h, \, \lambda'_N \lambda_N} \,&=&\, 
\frac{e^{2}}{Q^{2}} \, \bar{u}(k', h)\gamma _{\mu }u(k, h)\, \\
&\times& \, 
\bar{u}(p', \lambda'_N)\left( \tilde{G}_{M}\, \gamma ^{\mu }
-\tilde{F}_{2}\frac{P^{\mu }}{M}
+\tilde{F}_{3}\frac{\gamma\cdot K P^{\mu }}{M^{2}}\right) u(p, \lambda_N) \ . \nonumber
\ea
Knowing both expansions of the scattering amplitude is useful, particularly when making comparison to other work.  Our analysis will primarily use the second expansion, with the invariants denoted with tildes.  A selection of expressions using the primed invariants is given in Appendix~\ref{sec:axial}.

The scalar quantities $\tilde{G}_{M}$, $\tilde{F}_{2}$, and
$\tilde{F}_{3}$  are complex functions of two variables, say $\nu$ and
$Q^{2}$.  We will also use
\ba
\tilde{G}_{E}\equiv\tilde{G}_{M}-(1+\tau )\tilde{F}_{2}  \ .
\ea

In order to easily identify the one- and two-photon exchange contributions, we
introduce the notation $\tilde G_M = G_M + \delta \tilde G_M$, and
$\tilde G_E = G_E + \delta \tilde G_E$, where $G_M$ and $G_E$ are the
usual proton magnetic and electric form factors, which are functions of
$Q^2$ only and are defined from matrix elements of the electromagnetic
current. The amplitudes $ \tilde{F}_{3} = \delta \tilde F_3$, $\delta
\tilde{G}_{M}$, and $\delta \tilde G_E$, originate from processes
involving the exchange of at least two photons, and are of order $e^2$
(relative to the factor \( e^{2} \) in Eq.~(\ref{eq:tmatrix})).

The cross section without polarization is
\ba													\label{eq:reduced}
{d\sigma \over d\Omega_{Lab}} = 
				{ \tau \sigma_R  \over \epsilon(1+\tau)}
				{d\sigma_{NS} \over d\Omega_{Lab}}  \ ,
\ea

\noindent  where $\tau \equiv Q^2 / (4 M^2)$,  
$\varepsilon$ is
\ba
\varepsilon = \left( 1 + 2(1+\tau) \tan^2 {\theta\over 2} \right)^{-1}
		= {(s-u)^2 + t (4M^2 - t) \over (s-u)^2 - t (4M^2 - t) } \ ,
\ea

\noindent  $\theta$ is the electron Lab scattering angle, the ``no structure'' cross section is
\ba
{d\sigma_{NS} \over d\Omega_{Lab}} 
		= {4\alpha^2 \cos^2{\theta\over 2} \over Q^4} \,
		{E^{\prime 3} \over E}  \ ,
\ea

\noindent and $E$ and $E'$ are the incoming and outgoing electron Lab
energies.  For one-photon exchange, $\varepsilon$ is the polarization parameter of the virtual photon.  The reduced cross section including the two-photon exchange
correction is given by~\cite{GV03}
\be
\sigma_R 
= G_M^2 + \frac{\varepsilon}{\tau}  G_E^2   \,  
+ 2 \, G_M {\cal R} 
\left(\delta \tilde G_M + \varepsilon \frac{\nu}{M^2} \tilde F_3 \right) 
+ 2 \frac{\varepsilon}{\tau} G_E \, 
{\cal R} \left(\delta \tilde G_E + \frac{\nu}{M^2} \tilde F_3 \right) 
+  {\mathcal{O}}(e^4) ,
\label{eq:crossen} 
\ee
where \( {\cal R} \) stands for the real part.  Comparison to results elsewhere is often facilitated by the expression
\ba
{\nu \over M^2} = {s-u \over 4 M^2} = 
	\sqrt{ \tau (1+\tau) \frac{1+\varepsilon}{1-\varepsilon} } \ .
\ea

The general expressions for the double polarization observables for an
electron beam of positive helicity ($ h = + 1/2$) and for a recoil
proton polarization along its momentum ($P_l$) or perpendicular, but in
the scattering plane, to its momentum ($P_s$) can be derived as (for
$m_e = 0$)~\cite{GV03}:
\begin{eqnarray}
P_s &=& A_s =										\\
&=&	-\,\sqrt{\frac{2\varepsilon (1 -\varepsilon)}{\tau}} \,\frac{1}{\sigma_R} \,
	\left\{G_E G_M 
	+ G_E \, {\cal R} \left(\delta \tilde G_M \right) 
	+ G_M \, {\cal R} \left(\delta \tilde G_E + \frac{\nu}{M^2} \tilde F_3
	\right) 
	+  {\mathcal{O}}(e^4) \right\}  ,
										\nonumber
																\\[1.75ex]
P_l &=& - A_l 	 = \sqrt{1 - \varepsilon^2}  \,\frac{1}{\sigma_R} \,
	\left\{G_M^2 
	+ 2 \, G_M \, {\cal R} \left(\delta \tilde G_M 
	+ \frac{\varepsilon}{1 + \varepsilon} \frac{\nu}{M^2} \tilde F_3 \right) 
	+  {\mathcal{O}}(e^4) \right\}, 
													     \nonumber 
\end{eqnarray}

\noindent  The polarizations are related to the analyzing powers $A_s$ or $A_l$ by time-reversal invariance, as indicated above.  Note that $P_l$ is precisely unity in the backward direction, $\varepsilon =0$.  This follows generally from lepton helicity conservation and angular momentum conservation.

An observable which is directly proportional to the two- (or multi-)
photon exchange is a single-spin observable which is given by the
elastic scattering of an unpolarized electron on a proton target
polarized normal to the scattering plane (or the recoil polarization
$P_n$ normal to the scattering plane, which is exactly the same
assuming time-reversal invariance). The corresponding single-spin
asymmetry, which we refer to as the target (or recoil) normal spin
asymmetry ($A_n$), is related to the absorptive part of the elastic $e
N$ scattering amplitude \cite{RKR71}. Since the one-photon exchange
amplitude is purely real, the leading contribution to $A_n$ is of order
$O(e^2)$, and is due to an interference between one- and two-photon
exchange amplitudes. The general expression for $A_n$ in terms of 
the
invariants for electron-nucleon elastic scattering is given by (in the
limit $m_e = 0$),
\begin{eqnarray}
						  					\label{eq:tnsa}
P_n &=& A_n =  			\\
&=& \sqrt{\frac{2 \, \varepsilon \, (1+\varepsilon )}{\tau}} \,\,
	\frac{1}{\sigma_R} 
	\left\{ - \, G_M \, {\cal I} 
	\left(\delta \tilde G_E + \frac{\nu}{M^2} \tilde F_3 \right) 
	\,+ \, G_E \, {\cal I} \left(\delta \tilde G_M 
	+ \left( \frac{2 \varepsilon}{1 + \varepsilon} \right) 
	\frac{\nu}{M^2} \tilde F_3 \right) \right\} 
	       \, ,  \nonumber 
\end{eqnarray}

\noindent    where $\cal I$ denotes the imaginary part.

Another single-spin observable is the normal beam asymmetry,  which is discussed elsewhere~\cite{normal_beam} and which is also zero in the one-photon exchange approximation.  It is proportional to the electron mass, and the asymmetry is of ${\cal O}(10^{-6})$ for GeV electrons.  It is possibly observable in low energy elastic muon-proton scattering.  It was measured in experiments at MIT/Bates and MAMI \cite{BSSA-Exp} at electron beam energies below 1 GeV. It is possibly also observable in low-energy elastic muon-proton scattering.


\section{The two-photon exchange contribution to elastic electron-quark scattering}

\label{sec:3}

\begin{figure}
\includegraphics[height=4.5cm]{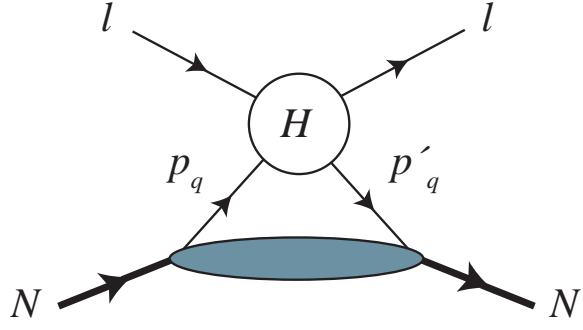}
\caption{Handbag approximation for the elastic lepton-nucleon
scattering at large momentum transfers. In the partonic scattering process 
(indicated by $H$), the lepton scatters from quarks in the nucleon, 
with momenta $p_q$ and $p'_q$. 
The lower blob represents the GPD's of the nucleon.}
\label{fig:handbag}
\end{figure}

In order to estimate the two-photon exchange contribution 
to $\tilde G_M$, $\tilde F_2$ and $\tilde F_3$ at large momentum
transfers, we will consider a partonic calculation illustrated in
Fig.~\ref{fig:handbag}. 
To begin, we calculate the subprocess on a quark, denoted 
by the scattering amplitude $H$ in Fig.~\ref{fig:handbag}. 
Subsequently, we shall embed the quarks in the proton as 
described through the nucleon's generalized parton distributions (GPD's).

\indent
Elastic lepton-quark scattering,
\begin{equation}
\label{eq:eqscatt}
l(k)+q(p_q)\rightarrow l(k')+q(p'_q) \ ,
\end{equation}
is described by two independent kinematical invariants, $\hat s \equiv
(k + p_q)^2$ and $Q^2 = -t = -(k - k')^2$. We also introduce the
crossing variable $\hat u \equiv (k - p'_q)^2$, which satisfies $\hat s
+ \hat u = Q^2$. The $T$-matrix for the two-photon part of the
electron-quark scattering can be written as
\begin{eqnarray}
\label{eq:tmatrixhard}
H_{h, \, \lambda} \,&=&\, 
\frac{(e \, e_q)^2}{Q^{2}} \, \bar{u}(k', h)\gamma _{\mu }u(k, h) \,\cdot \, 
\bar{u} (p'_q, \lambda) \left( \, \tilde{f}_{1} \, \gamma ^{\mu }
\, +\, \tilde{f}_{3} \, \gamma .K \, P_q^{\mu } \, \right) u(p_q, \lambda),
\end{eqnarray}
with $P_q \equiv (p_q + p'_q) / 2$, 
where $e_q$ is the fractional quark charge (for a flavor $q$), 
and where $u(p_q, \lambda)$ and $u(p'_q, \lambda)$ are the quark
spinors with quark helicity $\lambda = \pm 1/2$, which is
conserved in the scattering process for massless quarks.  Quark helicity conservation leads to the absence of any analog of $\tilde{F}_2$ 
in the general expansion of Eq.~(\ref{eq:tmatrix}).

\begin{figure}
\includegraphics[width=12.cm]{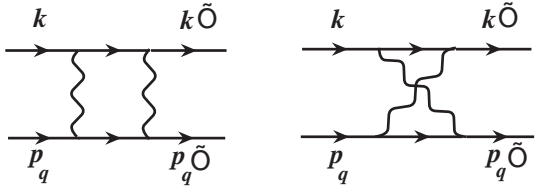}
\vspace{-6.cm}
\caption{Direct and crossed box diagrams to describe the 
two-photon exchange contribution to the lepton-quark scattering
process, corresponding with the blob denoted by $H$ in
Fig.~\ref{fig:handbag}.}
\label{fig:hardbox}
\end{figure}

In order to calculate the partonic scattering helicity amplitudes $H_{h, \lambda}$ of
Eq.~(\ref{eq:tmatrixhard}) at order $O(e^4)$, we consider the
two-photon exchange direct and crossed box diagrams of
Fig.~\ref{fig:hardbox}.  
%
%
The two-photon exchange contribution to the 
elastic electron-scattering off spin 1/2 Dirac particles was
first calculated in Ref.~\cite{Nie71}, which we verified explicitly. 
For further use, we separate the amplitude $\tilde f_1$ for the 
scattering of massless electrons off massless quarks 
into a soft and hard part, i.e. 
$\tilde{f}_1 = \tilde{f}_1^{soft} + \tilde{f}_1^{hard}$. 
The soft part corresponds with the situation where 
one of the photons in Fig.~\ref{fig:hardbox} carries zero four-momentum, 
and is obtained by replacing the other photon's four-momentum by 
$q$ in both numerator and denominator of the loop integral~\cite{grammer}. This yields,
\begin{eqnarray}
{\cal R}\left( \tilde{f}_1^{soft} \right)
&=&\frac{e^2}{4 \pi^2}\,
\left\{ \ln \left( \frac{\lambda^2}{\sqrt{- \hat s \hat u}} \right) 
\ln \left| \frac{\hat s}{\hat u} \right| + \frac{\pi^2}{2} \right\}, 
\label{eq:f1soft} \\
{\cal R}\left( \tilde{f}_1^{hard} \right) 
&=& \frac{e^2}{4 \pi^2} 
\left\{ \, \frac{1}{2} \, \ln \left| \frac{\hat s}{\hat u}  \right| 
+ \frac{Q^2}{4} 
\left[ \frac{1}{\hat u} \ln^2 \left| \frac{\hat s}{Q^2}  \right| 
- \frac{1}{\hat s} \ln^2 \left| \frac{\hat u}{Q^2} \right| 
- \frac{1}{\hat s} \pi^2  \right] \right\}, 
\label{eq:f1hard} 
\end{eqnarray}
where $\tilde f_1^{soft}$, which contains a term 
proportional to $\ln \lambda^2$ ($\lambda$ is 
an infinitesimal photon mass), is IR divergent.  
The amplitude $\tilde f_3$ resulting from the diagrams of 
Fig.~\ref{fig:hardbox} is IR finite, and its real part is
\be
{\cal R}\left( \tilde{f}_3 \right)
= \frac{e^2}{4 \pi^2} \, \frac{1}{\hat s \, \hat u} \,
\left\{ \hat s \, \ln \left| \frac{\hat s}{Q^2}  \right| + 
\hat u \, \ln \left| \frac{\hat u}{Q^2}  \right| 
+ \frac{\hat s - \hat u}{2} 
\left[ \frac{\hat s}{\hat u} \ln^2 \left| \frac{\hat s}{Q^2}  \right|  
- \frac{\hat u}{\hat s} \ln^2 \left| \frac{\hat u}{Q^2}  \right| 
- \frac{\hat u}{\hat s} \pi^2
\right] \right\} . 
\label{eq:f3q}
\ee

The correction to the electron-quark elastic cross section can be obtained from 
Eq.~(\ref{eq:crossen}),
\begin{eqnarray}
d \sigma \,&=&\, d \sigma_{1 \gamma} \, \left[ 
1 + 2 \, {\cal R} \left( \tilde{f}_1 \right)_{2 \gamma} + 
\varepsilon \, \frac{\hat s - \hat u}{4} \, 
 2 \, {\cal R} \left( \tilde{f}_3 \right)_{2 \gamma} \right] , \nonumber\\
&\equiv& d \sigma_{1 \gamma} \, \left( 1 + \delta_{2 \gamma} \right) ,  
\label{eq:crosseq}
\end{eqnarray}
where $d \sigma_{1 \gamma}$ is the cross section in the one-photon 
exchange approximation and  
$\varepsilon = - 2 \, \hat s \, \hat u \, / \, (\hat s^2 + \hat u^2)$ 
in the massless limit.  
Using Eqs.~(\ref{eq:f1soft}, \ref{eq:f1hard}, \ref{eq:f3q}), we obtain 
(for $e_q = +1$)
\begin{eqnarray}
\delta_{2 \gamma} &=& \frac{e^2}{4 \pi^2} \, 
\left\{ 2 \ln \left( \frac{\lambda^2}{Q^2}  \right) \, 
\ln \left| \frac{\hat s}{\hat u}  \right| \, \right. \\ \nonumber
&+& \left. \frac{(\hat s - \hat u) Q^2}{2 \, ( \hat s^2 + \hat u^2)}  
\left[ \ln^2 \left| \frac{\hat s}{Q^2}  \right| 
+ \ln^2 \left| \frac{\hat u}{Q^2} \right| 
+ \pi^2 \right] 
+   \frac{Q^4}{\hat s^2 + \hat u^2} \, 
\left[ \frac{\hat u}{Q^2} \ln \left| \frac{\hat s}{Q^2}  \right| 
- \frac{\hat s}{Q^2}\ln \left| \frac{\hat u}{Q^2} \right| \right] \, 
\right\} ,
\end{eqnarray}
which is in agreement with the corresponding expression for 
electron-muon scattering obtained in Ref.~\cite{Byt03}. 
The expressions of $\tilde f_1$ and $\tilde f_3$ can also 
be obtained through crossing from 
the corresponding expressions of the box diagrams for 
the process $e^+ e^- \to \mu^+ \mu^-$ as calculated in Ref.~\cite{Khr73}.

We will need the expressions for the 
imaginary parts of $\tilde f_1$ and $\tilde f_3$ in order to 
calculate the normal spin asymmetry $A_n$. These imaginary parts 
originate solely from the direct two-photon exchange box diagram 
of Fig.~\ref{fig:hardbox} and are given by
\begin{eqnarray}
{\cal I}\left( \tilde{f}_1^{soft} \right)
&=& - \frac{e^2}{4 \pi} \, 
\ln \left( \frac{\lambda^2}{\hat s} \right),  
\label{eq:f1qimsoft} 
											\\
{\cal I}\left( \tilde{f}_1^{hard} \right)
&=& - \frac{e^2}{4 \pi} 
\left\{ \frac{Q^2}{2 \, \hat u}  
\ln \left( \frac{\hat s}{Q^2}  \right) 
+ \frac{1}{2} \right\}, 
\label{eq:f1qimhard} 
											\\
{\cal I}\left( \tilde{f}_3 \right)
&=& - \frac{e^2}{4 \pi} 
\, \frac{1}{\hat u} \, \left\{ 
\frac{\hat s - \hat u}{\hat u}  
\ln \left( \frac{\hat s}{Q^2}  \right) 
+ 1 \right\}.  
											\label{eq:f3qim}
\end{eqnarray}
Notice that the IR divergent part in Eq.~(\ref{eq:f1qimsoft}) 
does not contribute when 
calculating the normal spin asymmetry $A_n$ 
of Eq.~(\ref{eq:tnsa}). 
Indeed at the quark level, one may complete the calculation for quark mass $m_q$ nonzero and find that  $A_n$ is given by,
\begin{eqnarray}
A_n =
\frac{e_q e^2}{4 \pi}  \frac{m_q}{2 \, Q} 
\, 
	\frac{\sqrt{2 \varepsilon (1 + \varepsilon)}}{1+4 \varepsilon m_q^2/ Q^2 }
\, 
		\frac{Q^2 (Q^2 + 4m_q^2)}{ \hat s(\hat s - \hat u)} 
\, 
\,   
\label{eq:anq} 
\end{eqnarray}
an IR finite quantity ({\it cf.} \cite{Barut60}). 


\section{The handbag calculation of the two-photon exchange contribution to 
elastic electron-nucleon scattering}

\label{sec:4}


Having calculated the partonic subprocess, 
we next discuss how to embed the quarks in the nucleon. 
We begin by discussing the soft contributions. The handbag diagrams discussed so far have both photons coupled to the same quark.  There are also 
contributions from processes where the photons interact with different quarks. 
One can show that the IR contributions from these processes, which are 
proportional to the products of the charges of the interacting quarks, 
added to the soft contributions from the handbag diagrams  
give the same result as the soft contributions calculated with just 
a nucleon intermediate state~\cite{Brodsky:1968ea}. Thus the low energy theorem for Compton scattering is satisfied. 
As discussed in the introduction, the hard parts which appear when the photons couple to different quarks, the so-called cat's ears diagrams, are neglected in the handbag approximation.

For the real parts, the IR divergence 
arising from the direct and crossed box diagrams, at the nucleon level,
is cancelled when adding the bremsstrahlung  
contribution  from the interference  of diagrams where a soft photon is emitted from the electron and from the proton. 
This provides a radiative correction term from the soft part of the boxes plus electron-proton bremsstrahlung 
which added to the lowest order term may be written as
\begin{eqnarray}
\label{eq:crosssoft}
\sigma_{soft} = \sigma_{1 \gamma} \, 
\left( 1 + \delta_{2 \gamma, \, soft} + \delta_{brems}^{e p} \right), 
\end{eqnarray}
where $\sigma_{1 \gamma}$ is the one-photon exchange cross 
section. In Eq.~(\ref{eq:crosssoft}), 
the soft-photon contribution due to the nucleon box diagram is given by
\begin{eqnarray}							
&&\delta_{2 \gamma , \, soft} 
= \frac{e^2}{2 \pi^2} \left\{ \ln \left( 
\frac{\lambda^2}{\sqrt{(s - M^2) |u - M^2|}}  \right) \, 
\ln \left| \frac{s - M^2}{u - M^2} \right| \right. 
													\label{eq:delta2gsoft}
													\\  \nonumber 
&&- \left. L\left( \frac{s - M^2}{s} \right) 
- \frac{1}{2} \ln^2\left( \frac{s - M^2}{s}\right) 
+ {\cal R} \left[ L\left( \frac{u - M^2}{u} \right) \right] 
+ \frac{1}{2} \ln^2\left( \frac{u - M^2}{u}\right) + \frac{\pi^2}{2} 
\right\} , 
\end{eqnarray}
where  $L$ is the Spence function defined by
%
\begin{equation}
L(z) = - \int_0^z \, dt \, \frac{\ln(1 - t)}{t} \ .
\end{equation}
The bremsstrahlung contribution where a 
soft photon is emitted from an electron and proton line 
({\it i.e.}, by cutting one of the (soft) photon lines in Fig.~\ref{fig:hardbox})  
was calculated in Ref.~\cite{MT00}, which we verified explicitly, and is for the case that the outgoing electron is detected,
\begin{eqnarray}
\delta_{brems}^{e p} 
&=& \frac{e^2}{2 \pi^2} 
\bigg\{ \ln \left( \frac{4 \, (\Delta E)^2 \, (s-M^2)^2}
{\lambda^2 \, y \, ( u-M^2 )^2} \right) \, 
\ln \left( \frac{ s-M^2 }{ M^2-u } \right)  
						\nonumber \\
&& \qquad + \  L\left(1 - \frac{1}{y} \, \frac{ s-M^2 }{ M^2-u } \right)
- L\left(1 - \frac{1}{y} \, \frac{ M^2-u }{ s-M^2 } \right)
\bigg\} , 
\label{eq:deltabremsep}
\end{eqnarray}
where  $\Delta E \equiv E_e^{\, ' el} - E_e^{\, '}$ 
is the difference of the measured outgoing electron {\it lab} energy 
($E_e^{\, '}$) from its elastic value ($E_e^{\, ' el}$), and
$y \equiv (\sqrt{\tau} + \sqrt{1 + \tau})^2$.
One indeed verifies that the sum of 
Eqs.~(\ref{eq:delta2gsoft},\ref{eq:deltabremsep}) is IR finite.
When comparing with elastic $ep$ cross section data, which are 
usually radiatively corrected using the procedure of Mo and Tsai, Ref.~\cite{MoTsai68}, 
we have to consider only the difference of our 
$\delta_{2 \gamma, \, soft} + \delta_{brems}^{ep}$ relative to the ${\cal O}(Z^2)$ part, in their notation, of the radiative correction in \cite{MoTsai68}. 
Except for the $\pi^2/2$ term in Eq.~(\ref{eq:delta2gsoft}), this difference was found to be below $10^{-3}$ for all kinematics considered in Fig.~\ref{fig:cross}.

Having discussed the two-photon exchange contribution on the nucleon
when one of the two photons is soft, we next discuss the contribution
which arises from the hard part (that is, neither photons soft) of the
partonic amplitude coming from the box diagrams. This part of the
amplitude is calculated, in the kinematical regime where $s $, $-u$,
and $Q^2$ are large compared to a hadronic scale ($s, -u, Q^2 >> M^2$),
as a convolution between a hard scattering electron-quark amplitude and
a soft matrix element on the nucleon. It is convenient to choose a
frame where $q^+ = 0$, as in~\cite{DY70}, where we introduce light-cone
variables $a^\pm \propto (a^0 \pm a^3)$ and choose the $z$-axis along
the direction of $P^3$ (so that $P$ has a large + component).  We use the symmetric frame, as in~\cite{Die99}, where the external momenta and $q$ are
\ba
k &=&   \left[ \eta P^+,  \frac{1}{P^+} {Q^2 \over 4\eta} ,
		\frac{1}{2} \vec q_\perp \right] \,,
								\nonumber \\
k' &=&   \left[ \eta P^+,  \frac{1}{P^+} {Q^2 \over 4\eta} ,
		 - \frac{1}{2} \vec q_\perp \right]  \,,
								\nonumber \\[1ex]
q &=&   \left[ 0, 0, \vec q_\perp \right]   \,,
								\nonumber \\[1ex]
p &=&   \left[ P^+, \frac{1}{P^+} \left( M^2 + \frac{Q^2}{4}  \right), 
			- \frac{1}{2} \vec q_\perp \right]   \,,
								\nonumber \\
p' &=&    \left[ P^+, \frac{1}{P^+} \left( M^2 + \frac{Q^2}{4}  \right),
			\frac{1}{2} \vec q_\perp \right]  \,.
\ea
Then,
\ba
s = \frac{(1 + \eta)^2}{4 \eta} \, Q^2 + (1+\eta) M^2   \,,    \qquad
u = - \frac{(1 - \eta)^2}{4 \eta} \, Q^2 + (1-\eta)M^2 \,,
\ea
one may check that $s+u=2M^2+Q^2$ and also solve for the lepton light-front momentum fractions, $\eta = k^+ / P^+ =
k^{\prime \, +} / P^+$, as
\begin{eqnarray}
\eta = \frac{1}{Q^2 + 4 \, M^2} 
\left[ s - u - 2 \, \sqrt{M^4 - s \, u} \, \right].
\end{eqnarray}
For comparison, forward scattering in the CM, $\theta_{\rm CM}=0^\circ$, matches to $\eta=0$ and backward scattering, $\theta_{\rm CM} = 180^\circ$, matches to $\eta = (s-M^2)/s$.

In the $q^+ = 0$ frame, the parton light-front momentum fractions are
defined as $x = p_q^+ / P^+ = p_q^{\prime \, +} / P^+$. The active
partons, on which the hard scattering takes place, are approximately
on-shell.   In the symmetric frame, we take the spectator partons to
have transverse momenta that are small (relative to $P$) and can be
neglected when evaluating the hard scattering amplitude $H$ in
Fig.~\ref{fig:handbag}. The Mandelstam variables for the process
(\ref{eq:eqscatt}) on the quark, which enter in the evaluation of the
hard scattering amplitude, are then given by
\begin{eqnarray} 
\label{eq:mandq}
\hat s = \frac{(x + \eta)^2}{4 \, x \, \eta} \, Q^2 \, , \hspace{1cm} 
\hat u = - \frac{(x - \eta)^2}{4 \, x \, \eta} \, Q^2 .
\end{eqnarray}
Note that in the limit $x \simeq 1$, where $\hat s \simeq s$ and $\hat
u \simeq u$ , the quark momenta are collinear with their parent hadron
momenta, i.e. $p_q \simeq p$ and $p'_q \simeq p'$. This is the simplest
situation for the handbag approximation, in which it was shown possible
to factorize the wide angle real Compton scattering amplitude in terms
of a hard scattering process and a soft overlap of hadronic light-cone
wave functions, which in turn can be expressed as moments of generalized
parton distributions (GPD's) \cite{Rad98,Die99}. In the following we
will extend the handbag~\cite{Die99,brodsky71} formalism to calculate the two-photon exchange
amplitude to elastic electron-nucleon scattering at moderately large
momentum transfers, and derive the amplitude within a more general
unfactorized framework by keeping the $x$ dependence in the hard
scattering amplitude ({\it i.e.}, by not taking the $x \to 1$ limit from the
outset).

For the process (\ref{Eq:intro.2}) in the kinematical regime 
$s, -u, Q^2 >> M^2$, the (unfactorized) handbag approximation implies
that the $T$-matrix can be written 
as\footnote{The corresponding equation in Ref.~\cite{YCC04} contains typographical errors regarding factors of $(1/2)$.  The remaining equations in that paper are written correctly.}
\begin{eqnarray}
\label{eq:handbag}
&& T_{h, \, \lambda'_N \lambda_N}^{hard} =
 \int_{-1}^1 \frac{dx}{x} \, \sum_q \frac{1}{2}
\left[ H_{h,\, + \frac{1}{2}}^{hard} + H_{h,\,- \frac{1}{2}}^{hard} \right] \times
										\\
&& \qquad \times  \frac{1}{2} \left[ H^q\left(x, 0, q^2 \right) \, 
\bar{u}(p', \lambda'_N) \, \gamma \cdot n \, u(p, \lambda_N) 
+ E^q\left(x, 0, q^2 \right)  \, 
\bar{u}(p', \lambda'_N) \, \frac{i \, \sigma^{\mu \nu} \, n_\mu q_\nu}{2 M} 
\, u(p, \lambda_N) \right] 
										\nonumber \\ \nonumber 
&& \qquad +  \int_{-1}^1 \frac{dx}{x} \, \sum_q \frac{1}{2}
\left[ H_{h,\, + \frac{1}{2}}^{hard} - H_{h,\,- \frac{1}{2}}^{hard} 
\right] \cdot \frac{1}{2} \mathrm{sgm}(x) 
\, \tilde H^q\left(x, 0, q^2 \right) 
\bar{u}(p', \lambda'_N) \, \gamma \cdot n \, \gamma_5 \, u(p, \lambda_N),
\end{eqnarray}
where the hard scattering amplitude $H^{hard}$ is evaluated 
using the hard part of $\tilde f_1$ and $\tilde f_3$, with kinematics 
$\hat s$ and $\hat u$ according to Eq.~(\ref{eq:mandq}), 
and where 
$n^\mu$ is a Sudakov four-vector ($n^2 = 0$), which can be expressed as
\begin{eqnarray}
n^\mu \,=\, \frac{2}{\sqrt{M^4 -  s u}} \, 
\left\{- \eta \, P^\mu \,+\, K^\mu \right\}.
\end{eqnarray}
Furthermore in Eq.~(\ref{eq:handbag}), 
$H^q, E^q, \tilde H^q$ are the GPD's for a quark $q$ in the
nucleon (for a review see, {\it e.g.}, Ref.~\cite{GPV01}).

>From Eqs.~(\ref{eq:tmatrix}), (\ref{eq:tmatrixhard}), and (\ref{eq:handbag})   
the hard $2 \gamma$ exchange contributions to 
$\delta \tilde G_M$, $\delta \tilde G_E$, and $\tilde F_3$ 
are obtained (after some algebra) as
\begin{eqnarray}
\hspace{-0.4cm}
\delta \tilde{G}_M^{hard} &=& C, 
\label{eq:GMhandbag}\\
\hspace{-0.4cm}
\delta \tilde{G}_E^{hard} &=& 
- \left( \frac{1 + \varepsilon}{2 \varepsilon} \right)\, ( A - C ) 
+ \sqrt{\frac{1 + \varepsilon}{2 \varepsilon}} \, B , 
\label{eq:GEhandbag} \\
\hspace{-0.4cm}
\tilde F_3 &=& \frac{M^2}{\nu}
\left( \frac{1 + \varepsilon}{2 \varepsilon} \right)\, (A - C ),
\label{eq:F3handbag}
\end{eqnarray}
with
\begin{eqnarray}
A &\equiv& \int_{-1}^1 \frac{dx}{x}     
\frac{\left[(\hat s - \hat u) \tilde{f}_1^{hard} -
\hat s \hat u \tilde{f}_3 \right]}{(s - u)} 
\sum_q e_q^2 \, \left( H^q + E^q \right), 
												\nonumber \\
B &\equiv& \int_{-1}^1 \frac{dx}{x}     
\frac{\left[(\hat s - \hat u) \tilde{f}_1^{hard} - 
\hat s \hat u \tilde{f}_3 \right]}{(s - u)} 
\sum_q e_q^2 \, \left( H^q - \tau E^q \right), 
												\nonumber \\
C &\equiv& \int_{-1}^1 \frac{dx}{x} \, \tilde{f}_1^{hard} \, 
\mathrm{sgm}(x) \, \sum_q e_q^2 \, \tilde H^q,
												\label{eq:ABC}
\end{eqnarray}
where note that in Eqs.~(\ref{eq:GMhandbag})-(\ref{eq:F3handbag}), 
the partonic amplitude $\tilde f_1$ has its soft IR divergent part 
removed as discussed before.

Equations~(\ref{eq:GMhandbag})-(\ref{eq:F3handbag}) reduce to the partonic amplitudes in the limit $M \to 0$ by considering a quark target for which the GPD's are given by,
\ba
H^q &\to& \delta(1-x) \,, \nonumber \\
E^q &\to& 0 \nonumber \,,  \\
\tilde H^q &\to& \delta(1-x) \,.
\ea
In this limit, and using the identity
\ba
- \frac{\hat s \hat u}{\hat s - \hat u} = \frac{\hat s - \hat u}{4}
			\frac{2\varepsilon}{1+\varepsilon} \,,
\ea
we find that
\ba
\delta \tilde G_M^{hard} &\to& \sum_q e^2_q \tilde f_1^{hard} \,, \nonumber \\
\frac{\delta \tilde F_2}{M} &\to& 0 \,, \nonumber \\
\frac{\delta \tilde F_3}{M^2} &\to& \sum_q e_q^2 \tilde f_3 \,.
\ea

>From the integrals $A$, $B$, and $C$, and the usual form factors,
we can directly construct the observables. 
The cross section is
\be
\sigma_R = \sigma_{R,soft} + \sigma_{R,hard} \,,
\ee
where
\ba
\sigma_{R,hard} 
	= (1 + \varepsilon) \, G_M \, {\cal R} \left( A \right)  
	+ \sqrt{2\, \varepsilon \, (1 + \varepsilon)} \frac{1}{\tau} \, 
			G_E \, {\cal R} \left( B \right)
	+ (1 - \varepsilon) \, G_M \, {\cal R} \left( C \right) \,.
\ea
>From Eqs.~(\ref{eq:crosssoft}) to~(\ref{eq:deltabremsep}) and the discussion surrounding them, we learned that to a good approximation the result for the soft part can be written as
\ba
\sigma_{R,soft} = \sigma_{R, 1\gamma} \left( 1 + \pi\alpha + \delta^{MT} \right) 
																		\,,
\ea
where $\delta^{MT}$ is the correction given in Ref.~\cite{MoTsai68}.  Since the data is very commonly corrected using~\cite{MoTsai68}, let us define
$\sigma_R^{\ MT\ corr} \equiv \sigma_R / (1+\delta^{MT})$.  Then an accurate relationship between the data with Mo-Tsai corrections already included and the form factors is
\ba
\sigma_R^{\ MT\ corr} 
	= \left( G_M^2 + \frac{\varepsilon}{\tau} G_E^2 \right)
		( 1 + \pi\alpha )
	+ \sigma_{R,hard} \,,
\ea
where the extra terms on the right-hand-side come from two-photon exchange and ${\cal O}(e^4)$ terms are not included.  The reader may marginally improve the expression by including with the $(1+\pi\alpha)$ factor the circa 0.1\% difference between our actual soft results and those of~\cite{MoTsai68};  from our side the relevant formulas are the aforementioned~(\ref{eq:crosssoft}) to~(\ref{eq:deltabremsep}).  Since the Mo-Tsai corrections are so commonly made in experimental papers before reporting the data, the ``$MT\ corr$'' superscript will be understood rather than explicit when we show cross section plots below.  Finally, before discussing polarization, the fact that a $\pi^2/2$ term, or $(\pi\alpha)$ term after multiplying in the overall factors, sits in the soft corrections has to do with the specific criterion we used, that of Ref.~\cite{grammer}, to separate the soft from hard parts.  The term cannot be eliminated; with a different criterion, however, that term can move into the hard part.

The double polarization observables of Eqs.~(\ref{eq:pt},\ref{eq:pl}) 
are given by
\begin{eqnarray}
P_s &=& -  
\,\sqrt{\frac{2\varepsilon (1 - \varepsilon)}{\tau}} \,\frac{1}{\sigma_R} \,
\left\{G_E G_M 
+ G_E \, {\cal R} \left( C \right) 
+ G_M \, \sqrt{\frac{1 + \varepsilon}{2 \, \varepsilon}} \, 
{\cal R} \left( B \right) 
+  {\mathcal{O}}(e^4) \right\}, 
\label{eq:pt} \\
P_l &=& 
\sqrt{1 - \varepsilon^2}  \,\frac{1}{\sigma_R} \,
\left\{G_M^2 
+ \, G_M \, {\cal R} \left( A + C \right) 
+  {\mathcal{O}}(e^4) \right\}, 
\label{eq:pl}
\end{eqnarray}
and the 
target normal spin asymmetry of Eq.~(\ref{eq:tnsa}) is
\begin{eqnarray}
A_n &=& \sqrt{\frac{2 \, \varepsilon \, (1+\varepsilon )}{\tau}} \,
\frac{1}{\sigma_R} \left\{
G_E \, {\cal I} \left( A \right) 
- \sqrt{\frac{1 + \varepsilon}{2 \varepsilon}} \, G_M \, 
{\cal I} \left( B \right)
\right\},
\label{eq:anhandbag}   
\end{eqnarray}
One sees from Eq.~(\ref{eq:anhandbag}) that $A_n$ does not depend on 
the GPD $\tilde H$.

We will need to specify a model for the GPD's in order to estimate the crucial integrals Eqs.~(\ref{eq:ABC}) for the two-photon exchange amplitudes  We will present results from two different GPD models: a gaussian model and a modified Regge model.  

First, following Ref.~\cite{Rad98}, we use a gaussian valence model which is unfactorized in $x$ and $Q^2$ for the GPD's $H$ and $\tilde H$,
\begin{eqnarray}
\label{eq:gpdh}
H^q(x, 0, q^2) \,=\, q_v(x) \; 
\exp \left(- \frac{(1 - x) \, Q^2}{4 \, x \, \sigma } \right) , \\
\label{eq:gpdht}
\tilde H^q(x, 0, q^2) \,=\, \Delta q_v(x) 
\; \exp \left(- \frac{(1 - x) \, Q^2}{4 \, x \, \sigma } 
\right) , 
\end{eqnarray}
where $q_v(x)$ is the valence quark distribution and $\Delta q_v(x)$ the 
polarized valence quark distribution. 
In the following estimates we take the unpolarized parton distributions 
at input scale $Q_0^2$ = 1 GeV$^2$ from the 
MRST2002 global NNLO fit~\cite{mrst02} as
\begin{eqnarray}
u_v &=& 0.262 \, x^{-0.69} (1 - x)^{3.50} 
\left( 1 + 3.83 \, x^{0.5} + 37.65 \, x \right), \nonumber \\
d_v &=& 0.061 \, x^{-0.65} (1 - x)^{4.03} 
\left( 1 + 49.05 \, x^{0.5} + 8.65 \, x \right) . \nonumber
\end{eqnarray}
For the polarized parton distributions, we adopt the recent NLO analysis of 
Ref.~\cite{Lea02}, which at input scale $Q_0^2$ = 1 GeV$^2$ yields
\begin{eqnarray}
\Delta u_v &=& 0.505  x^{-0.33} (1 - x)^{3.428} 
\left( 1 + 2.179  x^{0.5} + 14.57  x \right), \nonumber \\
\Delta d_v &=& - 0.0185  x^{-0.73} (1 - x)^{3.864} 
\left( 1 + 35.47  x^{0.5} + 28.97  x \right) . \nonumber
\end{eqnarray}
For the GPD $E$, whose forward limit is unknown, we adopt a valence 
parametrization multiplied with $(1 - x)^2$ 
to be consistent with the $x \to 1$ limit~\cite{Yuan03}. This gives
\begin{eqnarray}
\label{eq:gpde}
E^q(x, 0, q^2) = \frac{\kappa^q}{N^q} (1 - x)^2  q_v(x)  
\exp \left(- \frac{(1 - x) Q^2}{4 \, x \, \sigma } \right) ,
\end{eqnarray}
where the normalization factors $N^u=1.377$ and $N^d=0.7554$ are chosen in such a way that the first moments of $E^u$ and $E^d$ at $Q^2 = 0$ 
yield the anomalous magnetic moments 
$\kappa^u = 2 \kappa^p + \kappa^n = 1.673$ and 
$\kappa^d = \kappa^p + 2 \kappa^n = -2.033$ respectively.  
Furthermore, the parameter $\sigma$ in
Eqs.~(\ref{eq:gpdh},\ref{eq:gpdht},\ref{eq:gpde}) is related to the
average transverse momentum of the quarks inside the nucleon by 
$\sigma = 5 \, < k_\perp^2 > $.  Its value has been estimated in
Ref.~\cite{Die99} as $\sigma \simeq 0.8$~GeV$^2$, which we will adopt in the 
following calculations.

The GPD's just described were used in our shorter note~\cite{YCC04}.  Recently, GPD's whose first moments give a better account of the nucleon form factors have become available~\cite{guidal}.  These GPD's we refer to as a modified Regge model~\cite{guidal}, and entail
\ba
H^q(x, 0, q^2) &=& q_v(x) \; x^{a'_1 (1-x) Q^2}  \ ,
													\nonumber \\
E^q(x, 0, q^2) &=& \frac{\kappa^q}{N^q} (1-x)^{\eta_q}q_v(x) 
					\; x^{a'_2 (1-x) Q^2} \ ,  
													\nonumber \\
\tilde H^q(x, 0, q^2) &=& \Delta q_v(x) \; x^{ \tilde a'_1 (1-x) Q^2}   \ .
\ea
We still use the same $q_v$ and the same $\Delta q_v$ as given above.  The five parameters are
\ba
\alpha'_1 &=& 1.098 {\rm\ GeV}^{-2}, \qquad 
						\alpha'_2 = 1.158 {\rm\ GeV}^{-2}, \qquad
\tilde \alpha'_1 = 1.000 {\rm\ GeV}^{-2},
													\nonumber \\
\eta_u &=& 1.52, \qquad
						\eta_d = 0.31, 
\ea
and the normalization factors here become $N_u = 1.519$ and $N_d = 0.9447$.  The modified Regge GPD's formally do not give convergence at low $Q^2$ for integrands with negative powers of $x$, such as we have here (or as one finds in~\cite{Die99}).   The integrals could be defined by analytically continuing in the Regge intercept~\cite{Damashek:1969xj,Brodsky:1971zh}.  We will use them only for $Q^2 \ge 2$ GeV$^2$, and all the integrals converge straightforwardly.

We shall investigate in forthcoming plots the sensitivity of the results to the two GPD's.


\section{Results}

\label{sec:5}



\subsection{Cross section}

\begin{figure}

\includegraphics[width=7.8cm]{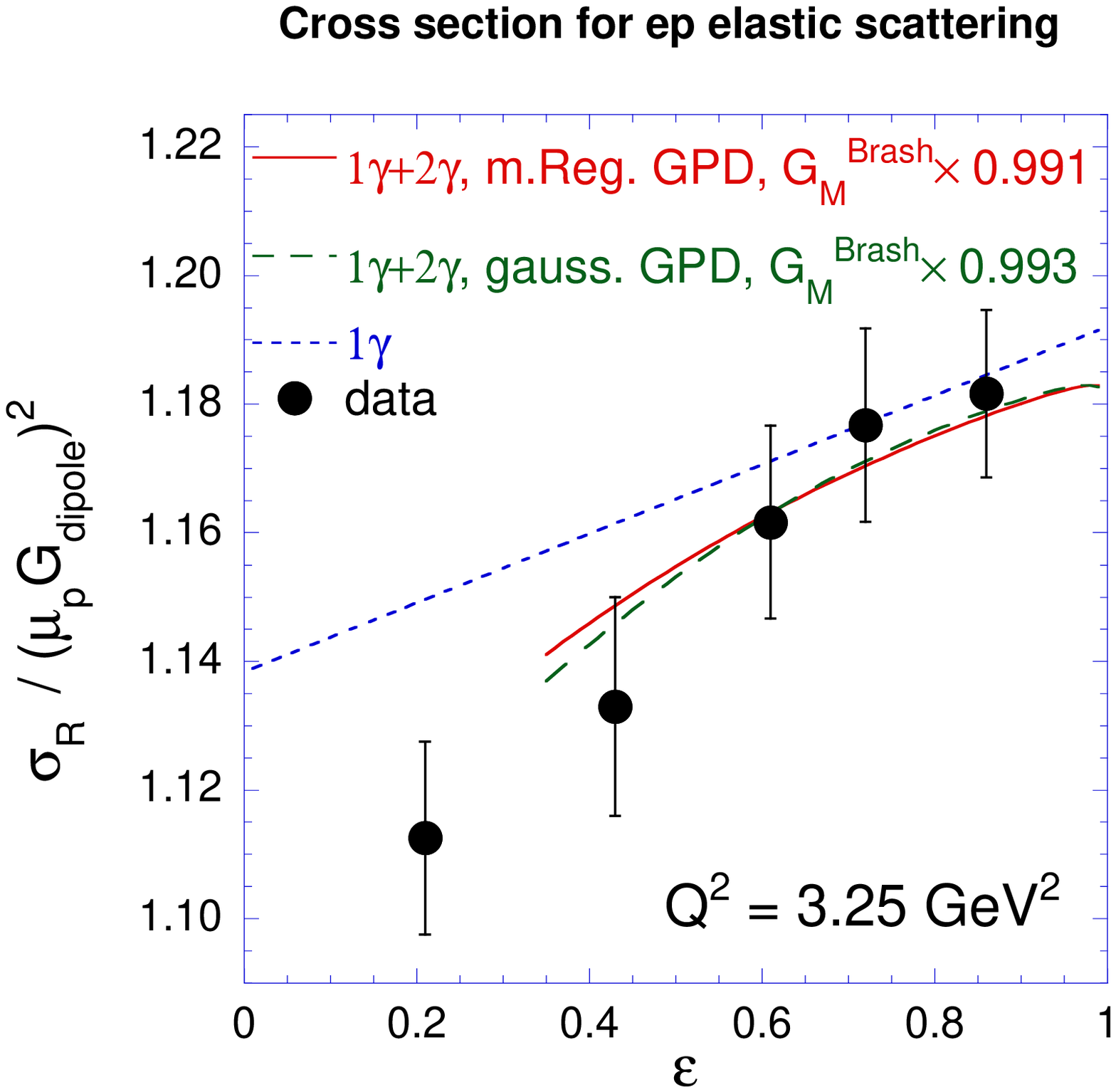} \hfill
\includegraphics[width=7.8cm]{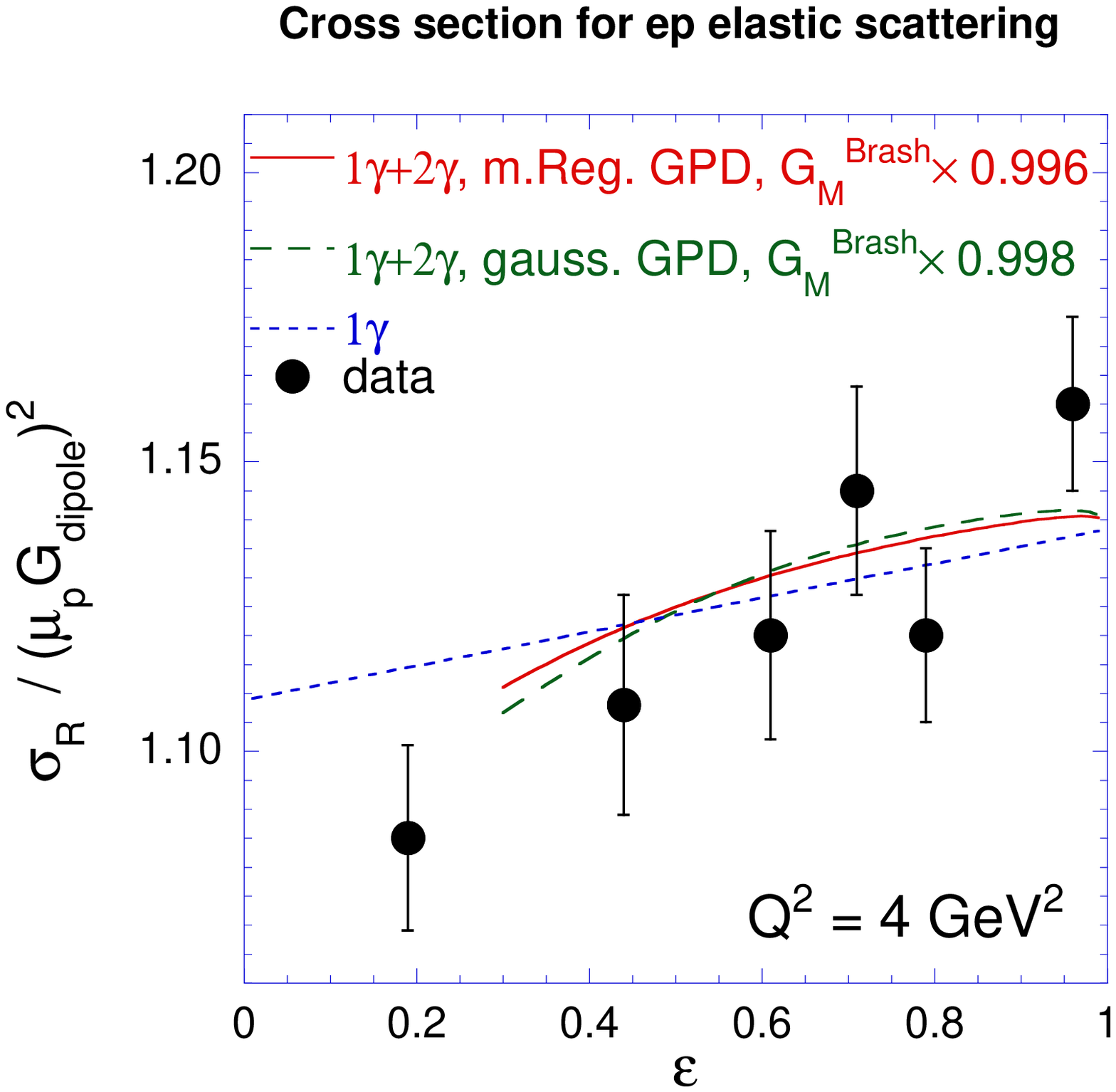} 

\vskip 5mm

\includegraphics[width=7.8cm]{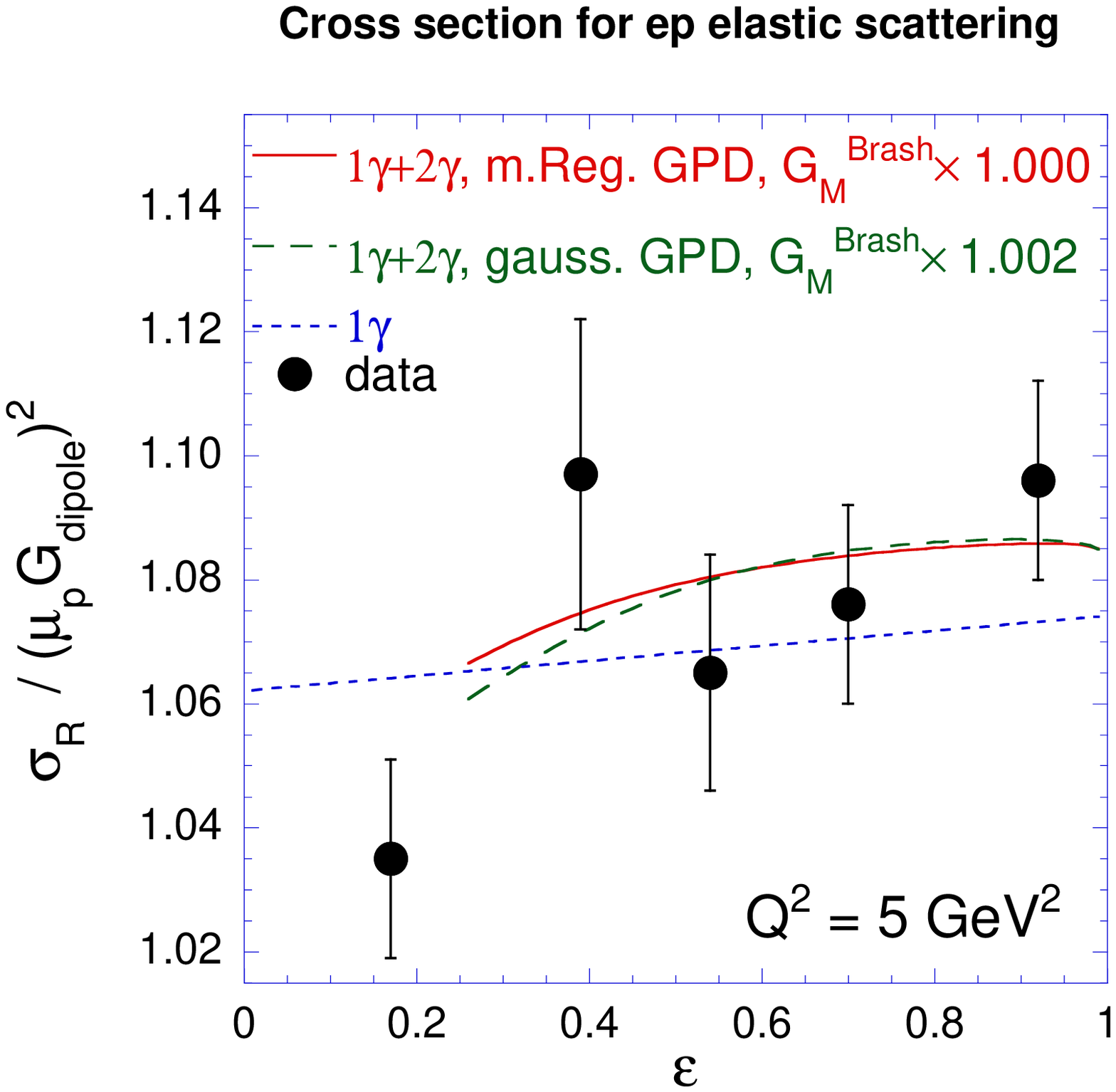} \hfill
\includegraphics[width=7.8cm]{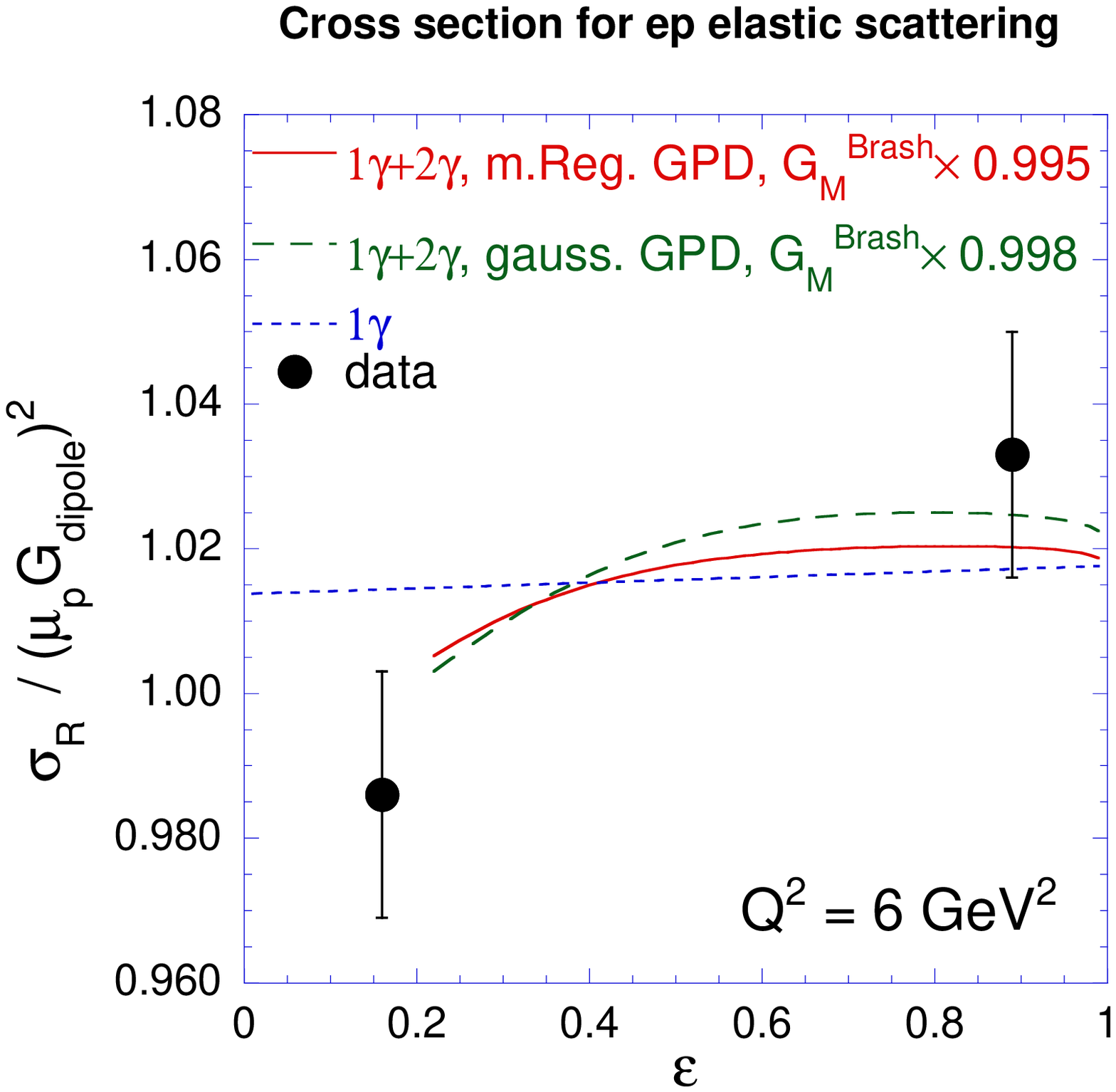}

\caption{Rosenbluth plots for elastic $e p $ scattering: 
$\sigma_R$ divided by $(\mu_p G_D)^2$, 
with $G_D = (1 + Q^2 / 0.71)^{-2}$.  
Dotted curves: Born approximation using $G_{E p} / G_{M p}$ from 
polarization data~\cite{Jones00,Gayou02}.  
Solid curves: full calculation using the modified Regge GPD, 
for the kinematical range $-u > M^2$. 
Dashed curves: same as solid curves but using the gaussian GPD.   
The data are from Ref.~\cite{Slac94}.}

\label{fig:cross}

\end{figure}

In Fig.~\ref{fig:cross}, we display the effect of $2 \gamma$ exchange
on the reduced cross section $\sigma_R$, as given in
Eq.~(\ref{eq:crossen}), for electron-proton scattering. For the form
factor ratio, we always use $G_E^p / G_M^p$ as extracted from the
polarization transfer experiments~\cite{Gayou02}.


We should remind the
reader that $G_M^p$ is also obtained from the reduced cross section
data: the normalization gives $G_M^p$ and the slope gives
$G_E^p/G_M^p$.  As a starting point we adopt the
parametrization for $G_M^{p}$, of Ref.~\cite{Bra02}.   
The straight dotted curves of Fig.~\ref{fig:cross} 
show that the  values of
$G_E^p/G_M^p$ extracted from the polarization experiments are
inconsistent with the one-photon exchange analysis of the Rosenbluth
data,  corrected with just the classic Mo and Tsai radiative
corrections \cite{MoTsai68}, in the $Q^2$ range where data from both methods exist.
We then include the $2 \gamma$ exchange correction, using the GPD based
calculation described in this paper.  The plots show the results from
both the GPD's used in our shorter note~\cite{YCC04} and recorded in the previous section, as well as from
the alternative GPD's also described in the previous section.  The results are
rather similar. 

It is also important to note the non-linearity in the
Rosenbluth plot, particularly at the largest
$\varepsilon$ values.   One sees that over most of the
$\varepsilon$ range, the overall slope has become steeper, in
agreement with the experimental data.  This change in slope is crucial:  we see that including the $2
\gamma$ exchange allows one to reconcile the polarization transfer and
Rosenbluth data.

 It is clearly worthwhile to do a global re-analysis of
all large $Q^2$ elastic data including the $2 \gamma$ exchange
correction in order
to redetermine the values of $G_E^p$ and $G_M^p$. 
For  example, in order to best fit the
data when including the $2 \gamma$ exchange correction,  one should 
slightly change the value of $G_M^p$ of Ref.~\cite{Bra02}. 
A full analysis is beyond the scope of this paper,


\begin{figure}

\includegraphics[width=7.8cm]{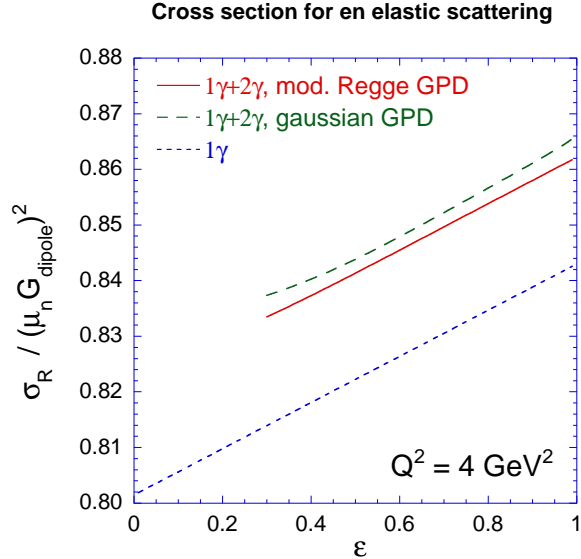} 

\caption{A Rosenbluth plot for elastic $e n$ scattering.  The curves with the two-photon corrections are plotted only for $-u>M^2$.  For $G_M^n$ and $G_E^n$ we used fits from~\cite{kubon} and~\cite{madey}, respectively. There is little slope change for the neutron case, for reasons noted in the text.}

\label{fig:cross-n}

\end{figure}


In Fig.~\ref{fig:cross-n}, we show a similar plot for electron-neutron elastic scattering.  Because of a partial cancellation between contributions proportional to $G_E^n$ and $G_M^n$, there is little $\varepsilon$ dependence in the corrections, and the slope is not appreciably modified.  We took $G_M^n$ from the fit of~\cite{kubon}; for $G_E^n$ we used the fit given in~\cite{madey}.


\subsection{Single spin asymmetry}

\begin{figure}

\includegraphics[height=7.5cm]{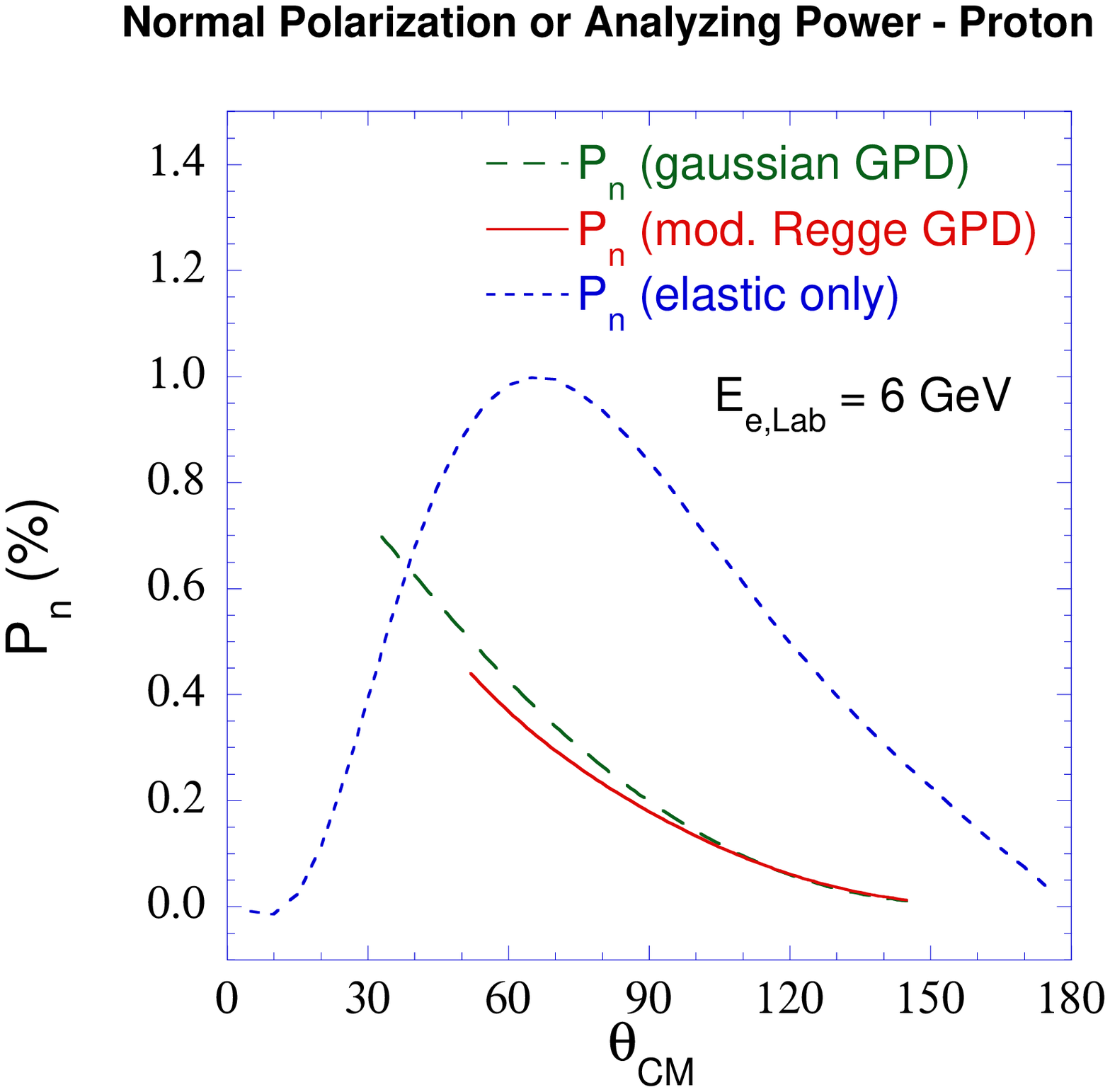}  \hfill
\includegraphics[height=7.5cm]{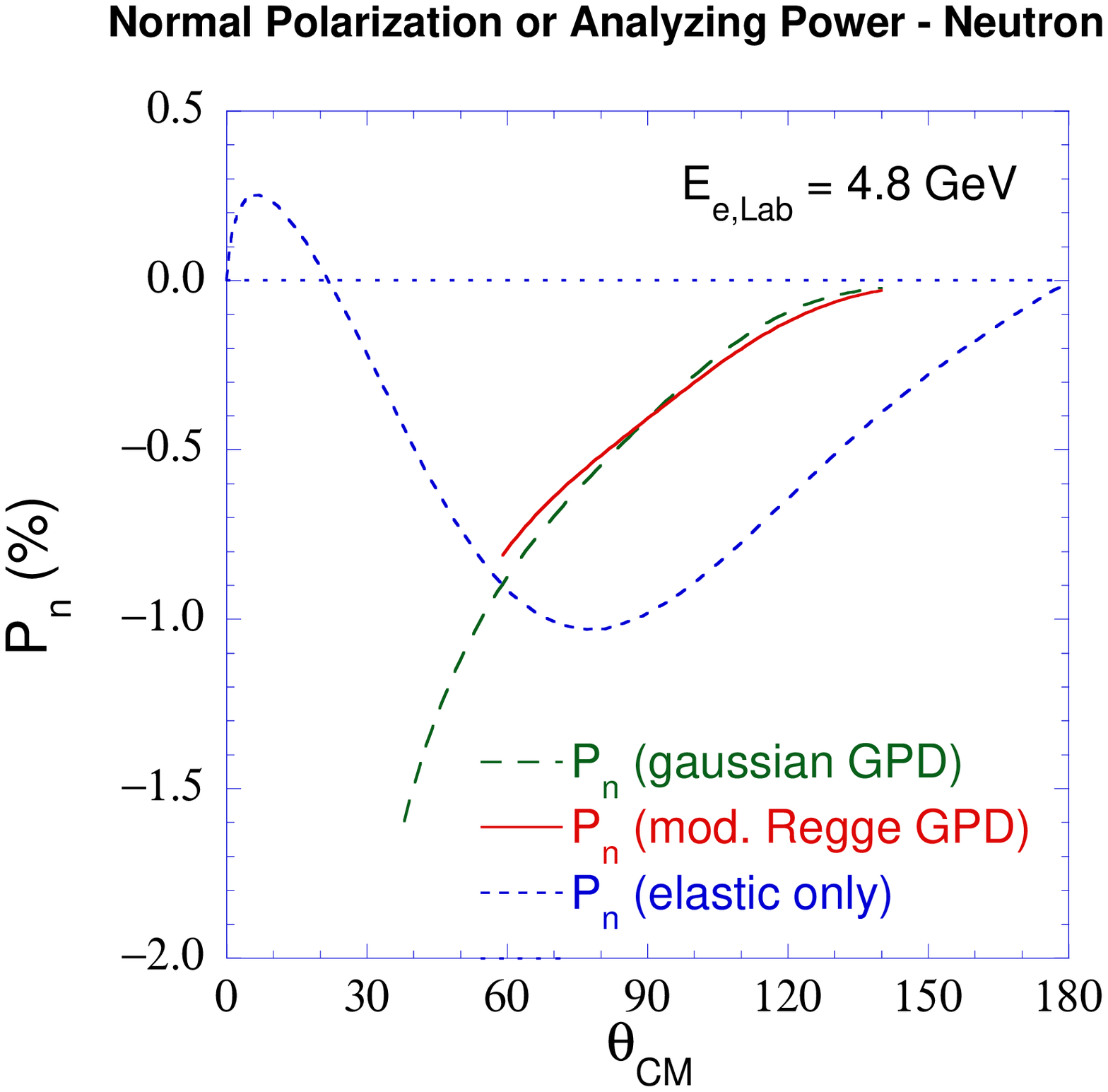}

\caption{Nucleon analyzing power, which is equal to normal recoil polarization.
The elastic contribution (nucleon intermediate state in the 
two-photon exchange box diagram) is shown by the dotted curve~\cite{RKR71}.
The GPD calculation for the inelastic contribution is 
shown by the dashed curve for the gaussian GPD, and by the solid curve for the modified Regge GPD.  The GPD calculation is cut off in the backward direction at $-u = M^2$.  
In the forward direction the modified Regge GPD result goes 
down to $Q^2 = 2$ GeV$^2$ and the gaussian GPD result to $Q^2 = M^2$.
}
\label{fig:ann}
\end{figure}

The single spin asymmetry $A_n$ or $P_n$ is a direct measure of the imaginary part of the $2 \gamma$ exchange amplitudes.  Our GPD estimate for $A_n$ for the proton is shown in the left-hand plot of Fig.~\ref{fig:ann} as a function of the CM scattering angle for fixed incoming electron lab energy,  taken here  as $6$ GeV.  Also shown is a calculation of $A_n$ including the elastic intermediate state only~\cite{RKR71}.  The result, which is nearly the same for either of the two GPD's that we use, is of order 1\%.

Fig.~\ref{fig:ann} on the right also shows a similar plot of the single spin asymmetry for a neutron target.  The predicted asymmetry is of opposite sign, reflecting that the numerically largest term is the one proportional to $G_M$.  The results are again of order 1\% in magnitude, though somewhat larger for the neutron than for the proton. 

A precision measurement of $A_n$ is planned at JLab \cite{AnTodd} on a polarized $^3He$ target; it will provide
access to the elastic electron-neutron single-spin asymmetry from two-photon exchange.


\subsection{Polarization transfers}



\begin{figure}
\includegraphics[width=7.8cm]{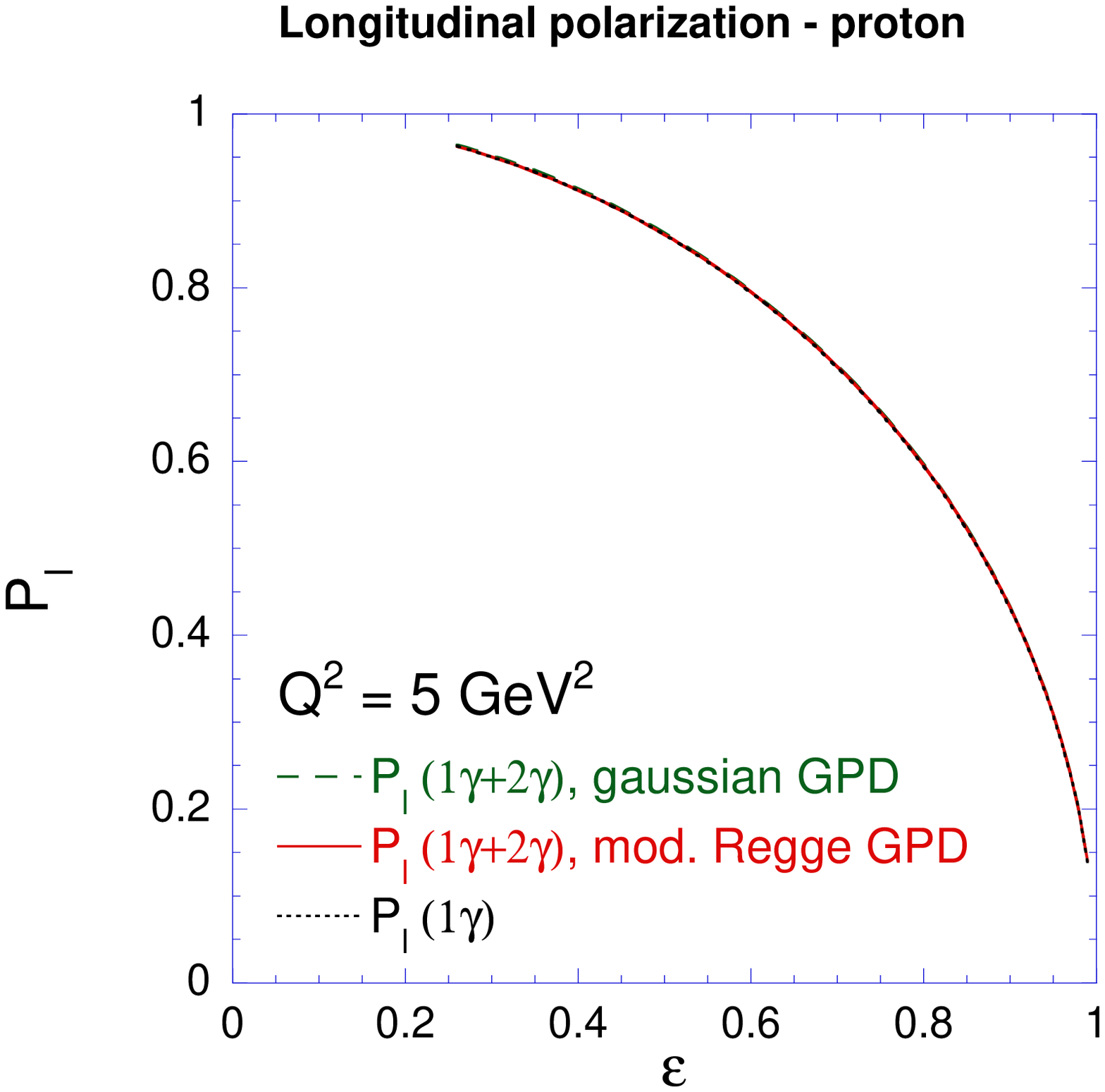} \hfill
\includegraphics[width=7.8cm]{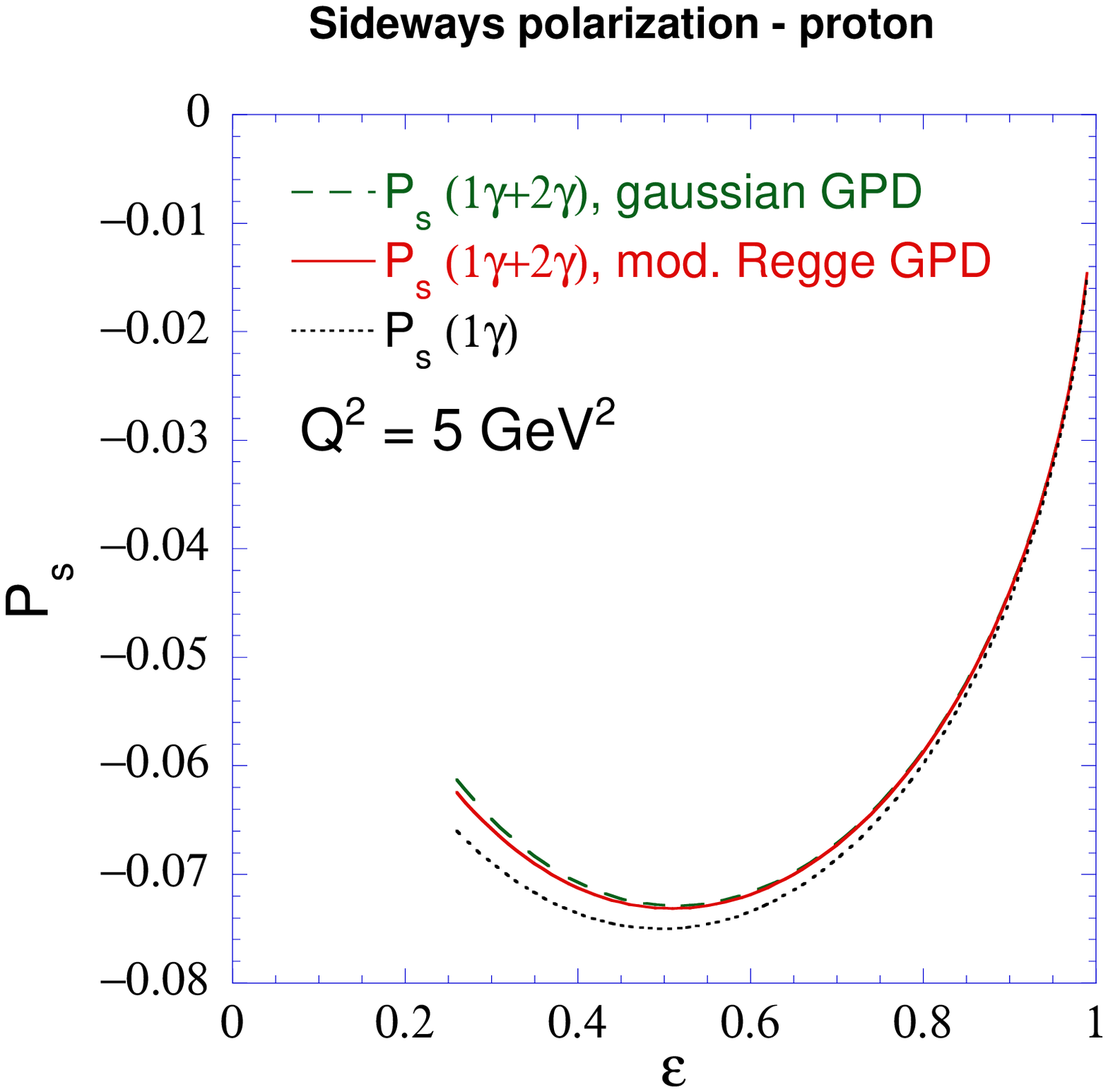}

\vskip 5mm

\includegraphics[width=7.8cm]{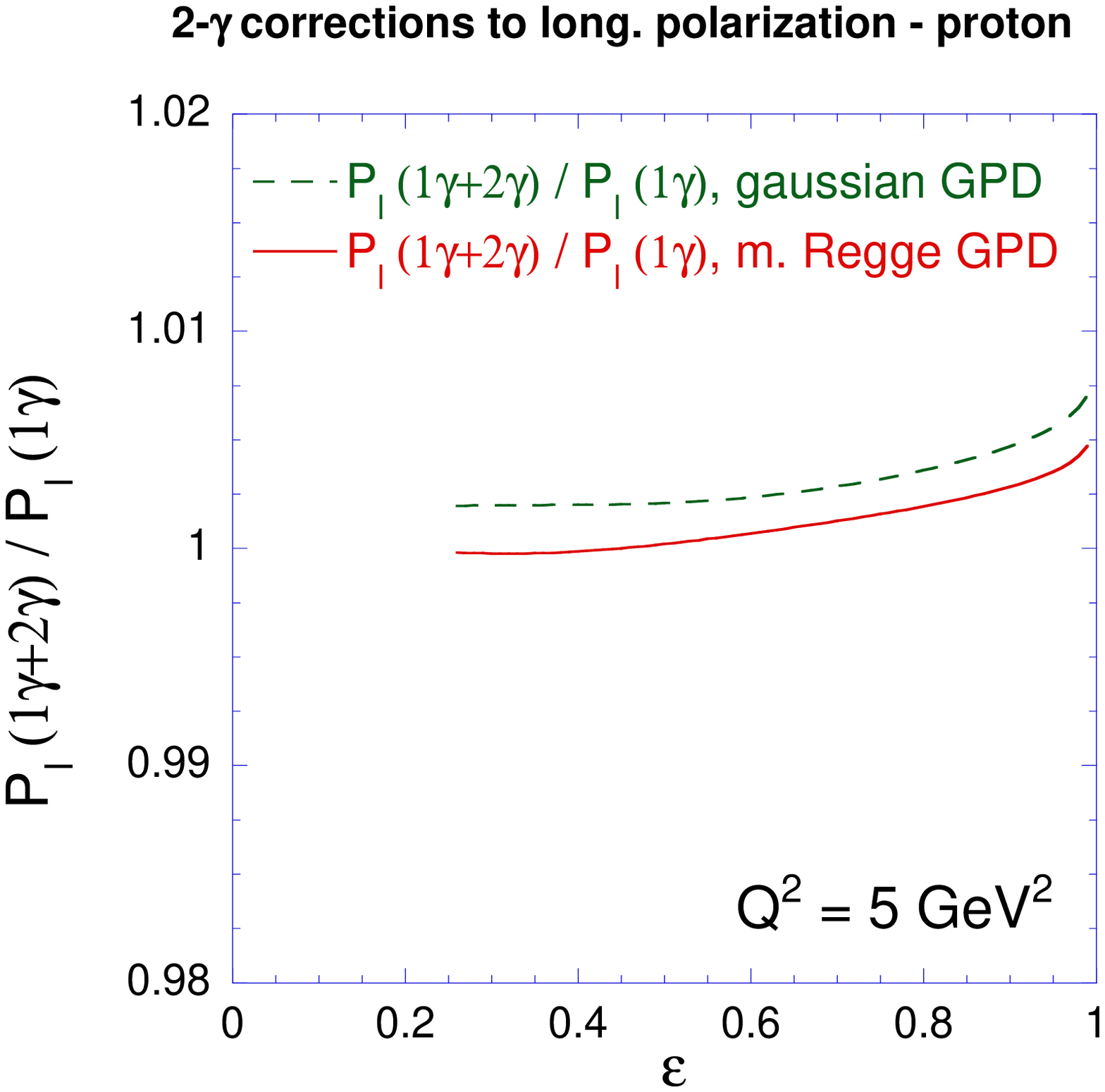} \hfill
\includegraphics[width=7.8cm]{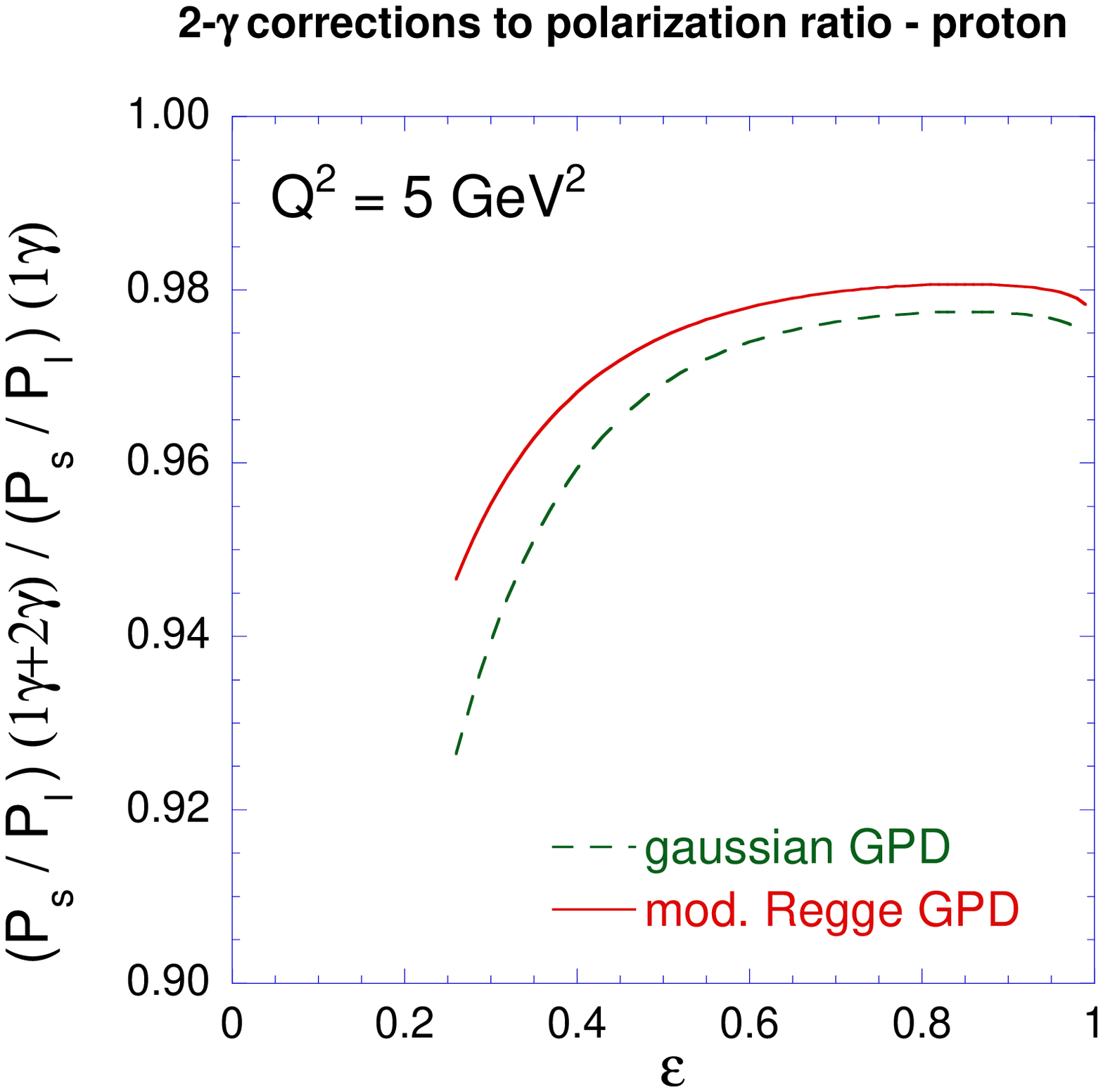}

\caption{Recoil proton polarization components $P_s$ and $P_l$ and their 
ratios relative to the $1 \gamma$ exchange results (lower panels)    
for elastic $e p$ scattering at $Q^2$ = 5 GeV$^2$. 
The dotted curves in the upper panels 
are the Born approximation ($1 \gamma$ exchange) results. 
The solid curves include the $2\gamma$ exchange correction using the 
GPD calculation, for the kinematical range where 
both $s, -u > M^2$.}
\label{fig:plpt}
\end{figure}


The polarization transfer method for measuring the ratio $G_E/G_M$ depends on measuring outgoing nucleon polarizations $P_l$ and $P_s$ for polarized incoming electrons.  Their ratio is
\be
{P_s \over P_l} = 
		- \sqrt{2\epsilon \over \tau (1+\epsilon)}\  
		{ G_E  \over G_M
		}  \ ,
\ee
in the one-photon exchange calculation.  This also is subject to additional corrections from two-photon exchange.  However, the impact of the corrections upon $G_E$ is not in any way enhanced, and so one expects and finds that the corrections to $G_E$ measured this way are smaller than the corrections to $G_E$ coming from the cross section experiments.

Figure~\ref{fig:plpt} shows in the upper two panels the calculated $P_l$ and $P_s$ for $ep$ scattering with and without the two-photon exchange terms, for 100\% right-handed electron polarization and with fixed momentum transfer $Q^2 = 5$ GeV$^2$.  The two GPD's were presented in the previous section, and we use again the polarization $G_E/G_M$ from~\cite{Gayou02} and $G_M$ from~\cite{Bra02}.  The corrections to the longitudinal polarization are quite small, as is seen again in the lower left panel,  where the ratio of the full calculation divided by the one-photon exchange calculation is shown.  The lower right panel shows the corrections to the $P_s / P_l$ ratio, given as a ratio again of the full calculation to the one-photon calculation.  An experiment to measure the $\varepsilon$-dependence of $P_s/P_l$ is planned at JLab~\cite{charles}.  This will allow a test of the two-photon corrections.


\begin{figure}
\includegraphics[width=7.8cm]{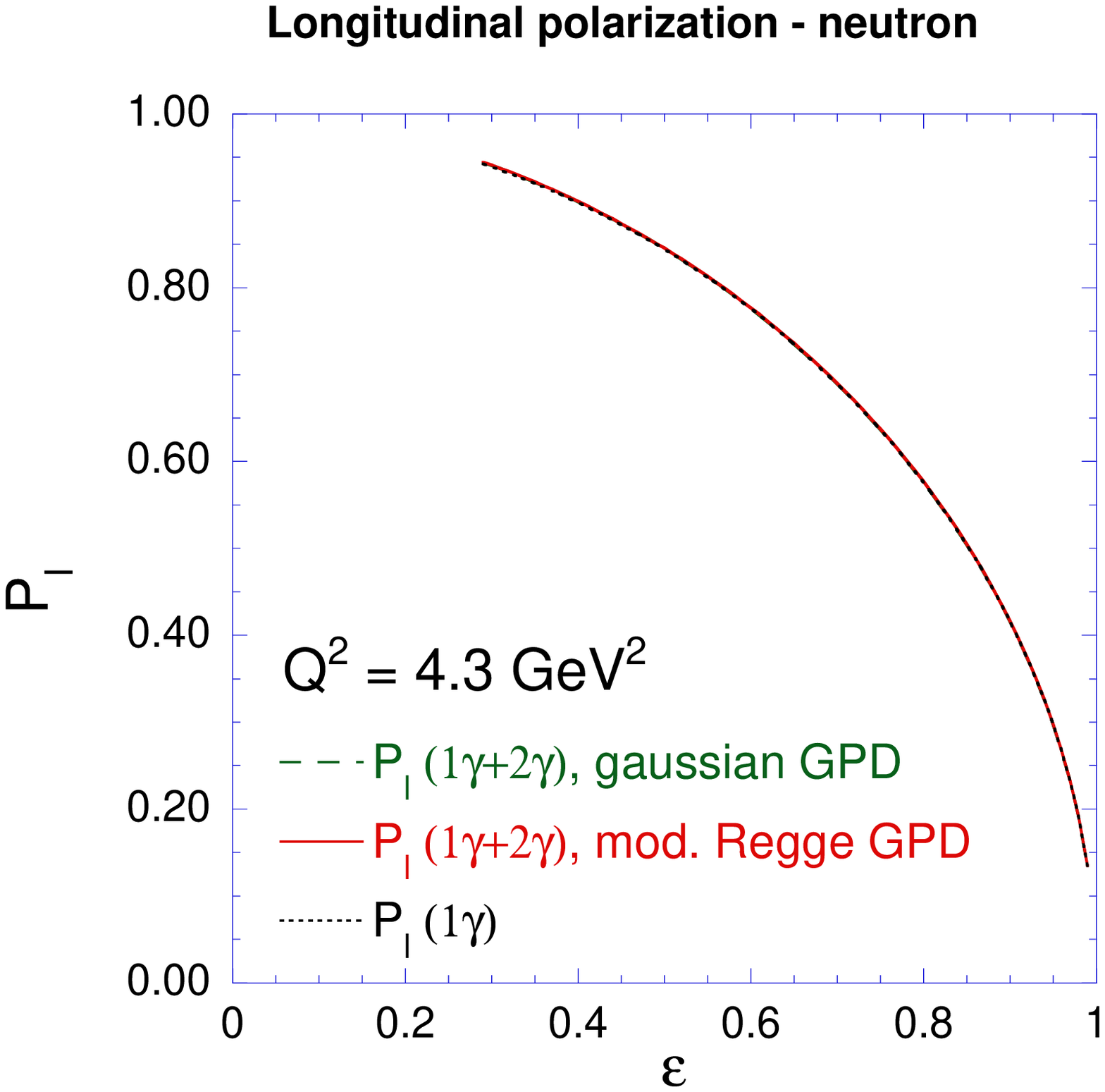} \hfill
\includegraphics[width=7.8cm]{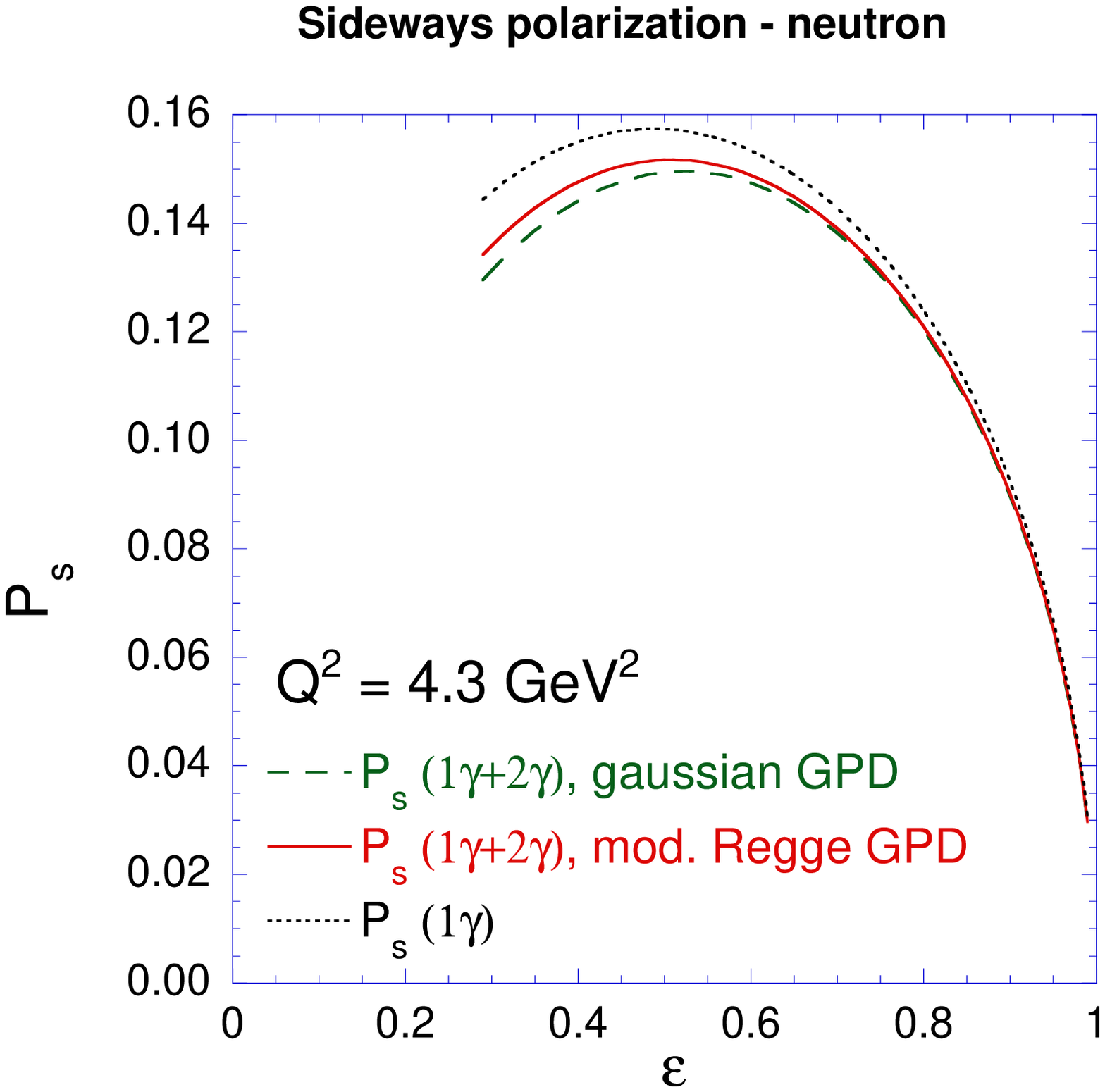}

\vskip 5mm

\includegraphics[width=7.8cm]{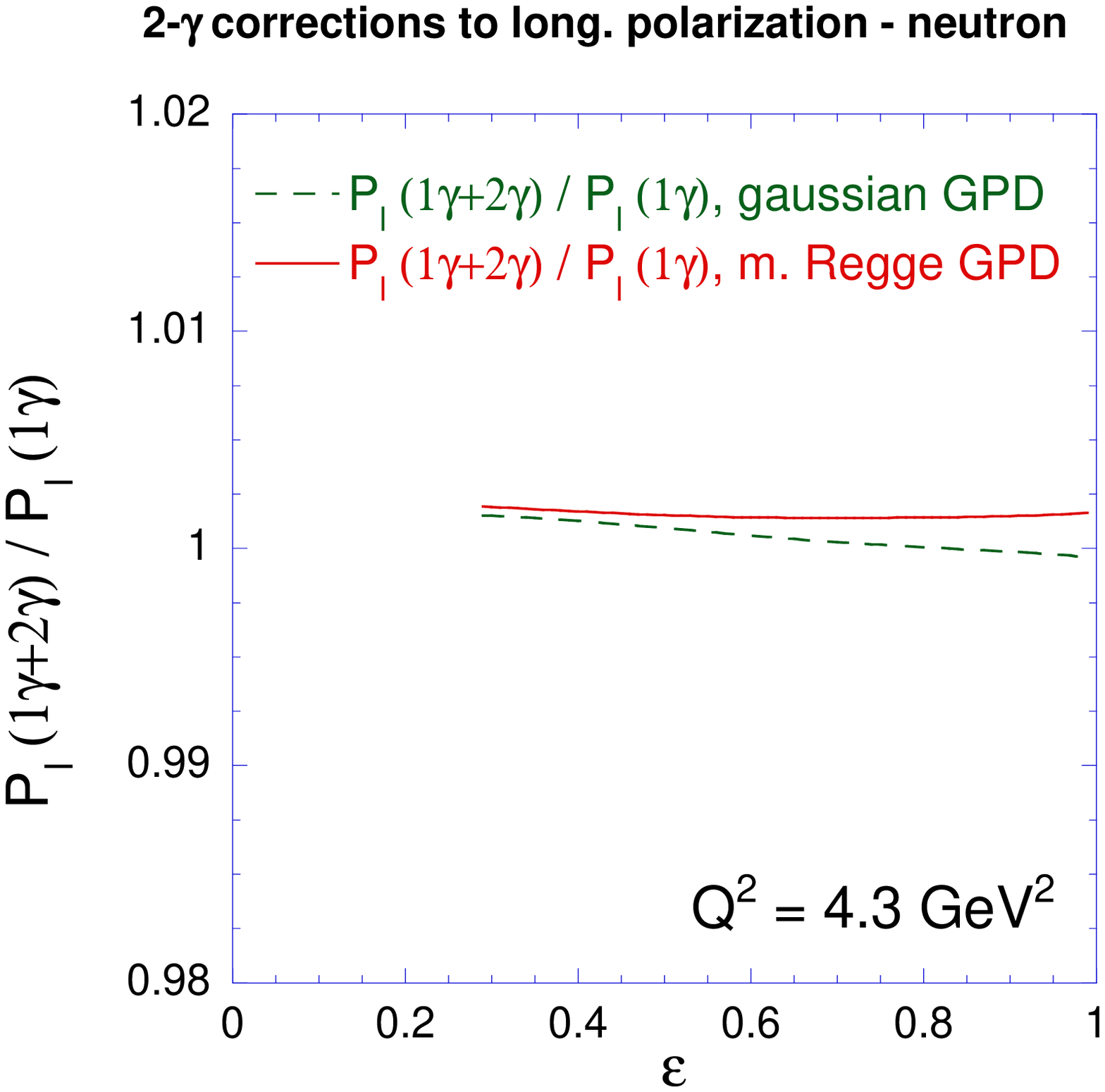} \hfill
\includegraphics[width=7.8cm]{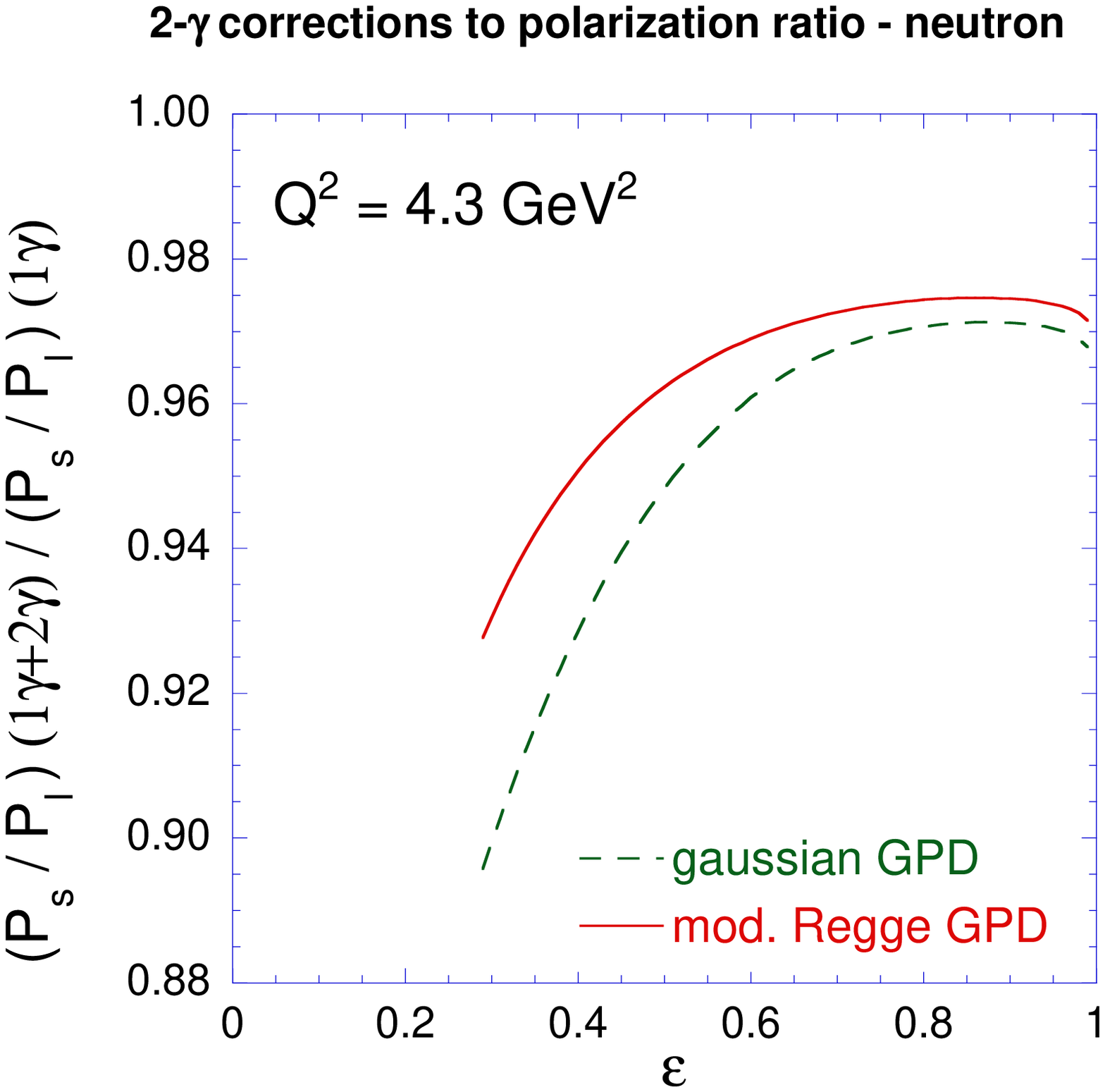}

\caption{Recoil neutron polarization components $P_s$ and $P_l$ and their 
ratios relative to the $1 \gamma$ exchange results (lower panels)    
for elastic $e p$ scattering at $Q^2$ = 4.3 GeV$^2$. 
See Fig.~\ref{fig:plpt} for notation.}
\label{fig:plpt-n}
\end{figure}


Figure~\ref{fig:plpt-n} shows the corresponding plots for the neutron, at a momentum transfer squared of 4.3 GeV$^2$.   If one needs to choose between the GPD's, the modified Regge model should be chosen as it gives the better account of the existing data on the form factors, the neutron form factors in particular~\cite{guidal}.


\subsection{Positron-proton vs. electron-proton}

\begin{figure}
\includegraphics[width=12.0cm]{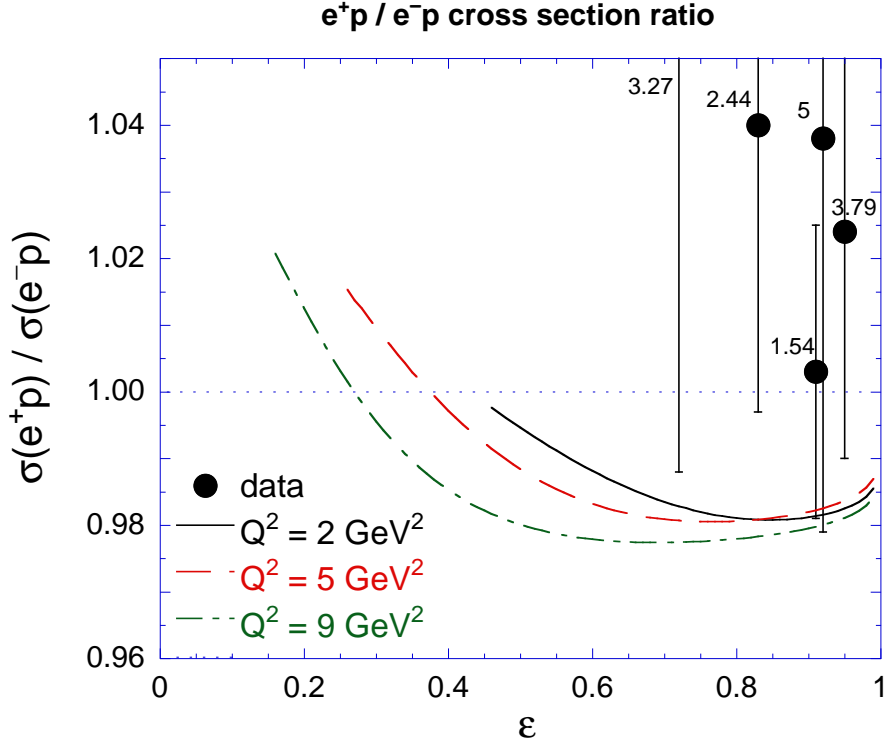}
\caption{Ratio of $e^+ / e^-$ elastic cross sections on the proton.    
The GPD calculations for the $2 \gamma$ exchange correction are for three fixed $Q^2$ values of 2, 5, and 9 GeV$^2$, 
for the kinematical range where $-u$ is above $M^2$. 
Also shown are all known data, from~\cite{Mar68}, with $Q^2$ above 1.5 GeV$^2$ (the missing central value is at 1.111).    The numbers near the data give $Q^2$ for that point in GeV$^2$.
}
\label{fig:positron}
\end{figure}

Positron-proton and electron-proton scattering have the opposite sign for the two-photon corrections 
relative to the one-photon terms.  Hence one expects $e^+ p$ and $e^- p$ elastic scattering to differ 
by a few percent.  Figure~\ref{fig:positron} shows our results for three different $Q^2$ values.  
These curves are obtained by adding our two-photon box calculation, minus the corresponding part of the soft 
only calculation in~\cite{MoTsai68}, to the one-photon calculations; hence, they are meant to be compared to 
data where the corrections given in~\cite{MoTsai68} have already been made.  Each curve is based on the gaussian 
GPD and is cut off at low $\varepsilon$ when $-u = M^2$.  Early data from SLAC are available~\cite{Mar68}; more 
precise data are anticipated from JLab~\cite{brooks}.  (Ref.~\cite{Mar68} used the Meister-Yennie~\cite{oldyennie} 
soft corrections rather than those of Mo and Tsai.  We have checked that for these kinematics the difference between 
them is  smaller than $ 0.1\%
,$  which is negligible compared to the size of the error bars.)


\subsection{Possibilities at lower $|u|$}   \label{subsectionF}


In numerical calculations, we used a conservative requirement that the
values of the Mandelstam variable $|u|>M^2$ in order to apply the
partonic description. In a `handbag' mechanism of wide-angle Compton
scattering on a proton, such a requirement is needed to enforce high
virtuality of the quark line between the two currents, making sure that short light-cone
distances dominate. However, our case of electron--quark scattering via
two-photon exchange involves 4-dimensional loop integration, and small
values of $|u|$ do not necessarily mean that the struck quark has small
virtuality. Analyzing the two-photon-exchange loop integral in terms of
Sudakov variables one may show that for the backward ($u\to0$)
electron--quark scattering, high virtuality of the quark dominates the
loop integral, thereby justifying  extension of our approach to the
region of small $u$, as long as  $s$ and $-t$ remain large. Such an
analysis may be found in the literature for the backward-angle
electron-muon scattering in QED~\cite{gglf}, and we found our formalism
consistent with these early calculations.


\subsection{Rosenbluth determinations of $G_E/G_M$ including  2-photon corrections}


Previous Rosenbluth determinations of $G_E/G_M$ were made using data which had been radiatively corrected using the Mo-Tsai~\cite{MoTsai68} or comparable~\cite{oldyennie} prescription.  Given the work in this paper, we would now say that these corrections are just a part of the total radiative correction.  One should also include the hard two-photon corrections.  

We present here new Rosenbluth determinations of $G_E^p/G_M^p$ using known data but including the two-photon corrections.    We used cross section data from Andivahis {\it et al.}~\cite{Slac94}, and made a $\chi^2$ fit to the data at each of the five $Q^2$ selected using our full calculation and allowing both $G_M^p$ and $G_E^p/G_M^p$ to vary.  We included the lowest $\varepsilon$ points in the data by making a linear extrapolation of our calculations from higher $\varepsilon$.   (For the record, and for the $\varepsilon$'s in question and to the precision we need, the result is numerically the same as doing our GPD calculation at these $\varepsilon$'s, even though $|u|$ is below $M^2$.)

\begin{figure}
\includegraphics[width=10.5cm]{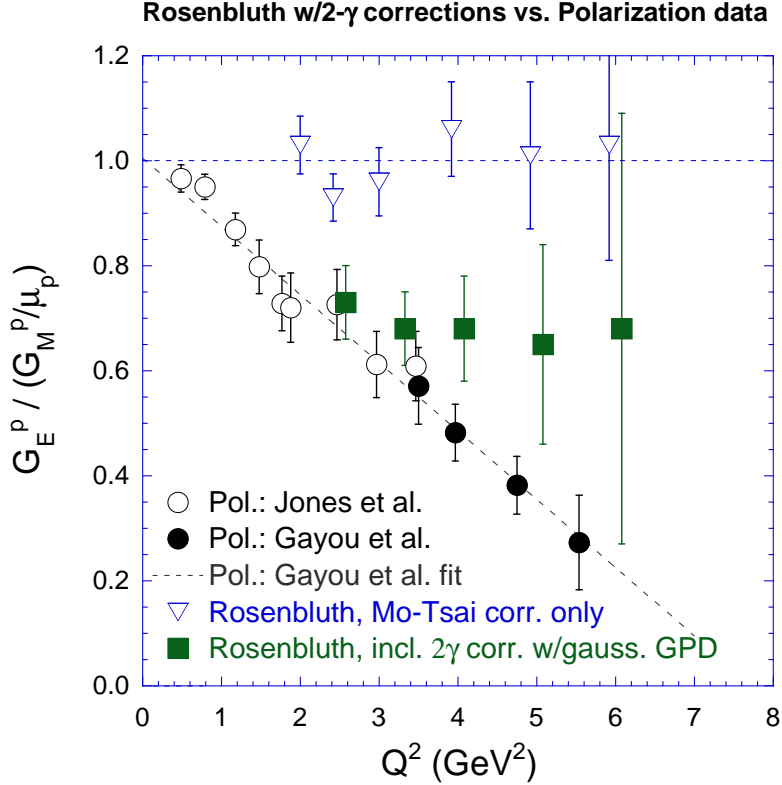}
\caption{Rosenbluth determinations of $G_E/G_M$ including the 2-photon corrections.  The polarization data is from Jones {\it et al.}~\cite{Jones00} and Gayou {\it et al.}~\cite{Gayou02}, and the Rosenbluth determinations without the two-photon corrections are from~\cite{Arrington:2003df}. The Rosenbluth $G_E/G_M$ are based on data from Andivahis {\it et al.}~\cite{Slac94}.  Some of our points for the Rosenbluth results are slightly offset horizontally for clarity.
}
\label{fig:rosencompare}
\end{figure}

The results are shown in Fig.~\ref{fig:rosencompare}.  The figure also shows the results of the polarization transfer measurements, and Rosenbluth results taken from~\cite{Arrington:2003df}, which do not include the hard two-photon corrections.  The polarization results also have radiative corrections, but the size of them is, as one has learned from Fig.~\ref{fig:plpt}, smaller than the dots of the data points.  The solid squares in Fig.~\ref{fig:rosencompare} show the $G_E/G_M$ ratios we have extracted with Ref.~\cite{Slac94} data and the two-photon corrections with the gaussian GPD.    The results with the modified Regge GPD are omitted to reduce clutter on the graph;  they are about the same as for the gaussian for $Q^2$ of 2--3 GeV$^2$, and a bit larger at the higher $Q^2$.

%


For $Q^2$ in the 2--3 GeV$^2$ range, the $G_E/G_M$ extracted using the Rosenbluth method including the two-photon corrections agree well with the polarization transfer results.  At higher $Q^2$, there is at least partial reconciliation between the two methods.   

One may comment on the growth of the error bars at higher $Q^2$.  The calculation with the two-photon contributions includes a lowest order term quadratic in $G_E^p$ and a correction linear in $G_E^p$ with opposite sign.  The partial cancellation explains the reduced sensitivity to changes in $G_E^p$.


\section{Conclusions}

\label{sec:theend}


We have studied the effects of two-photon physics  for lepton-nucleon elastic  scattering.   Our main result is a calculation of the two-photon exchange contributions including contributions coming when intermediate particles which are far off shell.  The main impediment to performing this calculation is the lack of knowledge of nucleon structure.  Here we have used a partonic ``handbag" model to express the contributions when both photons are hard in terms of the generalized parton distributions (GPD's) of the nucleon.  The GPD's also enter calculations of deeply virtual Compton scattering, wide angle Compton scattering, and exclusive meson photoproduction,  which are consistent with  models for the GPD's.   The calculations which we have presented are  valid when $s$, $-u$, and $Q^2$ are large, although we  have argued  in subsection~\ref{subsectionF} and Ref.~\cite{gglf}, that the requirement on $-u$ is not compulsory for $eN$ elastic scattering).   We have presented our results requiring that the magnitude of each of the invariants is above $M^2$.

We have found  that in Rosenbluth plots of the differential cross section vs. $\varepsilon$, that the two-photon exchange corrections gives an additional slope which is sufficient to reconcile qualitatively the difference between the Rosenbluth and polarization data.  The change in the effective slope in the Rosenbluth plots comes only from corrections where both photons are hard.  The reconciliation thus implies only a minor change in the $G_E/G_M$ ratio as obtained from the polarization data, since those data receive smaller two-photon corrections to $G_E/G_M$.

Two-photon exchange has additional consequences which could be experimentally observed.    For polarizations $P_s$ and $P_l$, there are two-photon  corrections which are small but measurable.   For the normal direction, the polarization or analyzing power  is zero in the one-photon exchange limit, but the presence of the two-photon exchange amplitude leads to a nonzero effect  of ${\cal O}(1\%
)$.  We also predict  a ${\cal O}({\rm few}\%
)$ positron-proton/electron-proton asymmetry.   The predicted Rosenbluth plot is no longer precisely linear; it acquires a measurable curvature, particularly at high $\varepsilon$.

Thus, in summary,  we have shown that the hard two-photon exchange mechanism substantially reconciles the Rosenbluth and polarization transfer measurements of the proton electromagnetic elastic form factors.  We have also emphasized that there are important experimentally testable consequences of the two-photon amplitude.

\section*{Acknowledgments}

We thank P. A. M. Guichon, N. Merenkov, and S.N. Yang for useful discussions.  This work was supported by the Taiwanese NSC under contract 
92-2112-M002-049 (Y.C.C.), 
by the NSF under grant PHY-0245056 (C.E.C.)
and by the U.S. DOE under contracts DE-AC05-84ER40150 (A.A., M.V.), DE-FG02-04ER41302 (M.V.), 
and DE-AC03-76SF00515 (S.J.B.). 




\appendix
\section{Cross section and polarization results in the axial-vector representation}

\label{sec:axial}


This appendix records the cross section and polarization results using the expansion of the scattering amplitude in the axial-vector representation given by Eq. (\ref{eq:alt}),
\begin{eqnarray}
T_{h, \, \lambda'_N \lambda_N} = {e^2 \over Q^2 }  &\Bigg\{&
		\bar u(k',h) \gamma_\mu u(k,h) \times 
		\bar u(p',\lambda'_N) \left[ \gamma^\mu G'_M - {P^\mu \over M} 		
			F'_2 \right] u(p,\lambda_N)
							\nonumber \\
		&+& \bar u(k',h) \gamma_\mu \gamma_5 u(k,h) \times 
		\bar u(p,\lambda'_N) \, \gamma^\mu \gamma^5 G'_A \,  u(p,\lambda_N)
						\ \ 	  	\Bigg\}  \ .
\end{eqnarray}


\subsection{Form factors and observables}


\noindent The scalar invariants or form factors are in general complex and functions of two variables.  We also define
\ba
G'_{E}\equiv G'_{M}-(1+\tau ) F'_{2}  \ .
\ea

\noindent The relations between the present scalar invariants and the ones used in most of the text follow from Eq.~(\ref{eq:theorem}) and are
\ba
G'_M &=& \tilde G_M + {s-u \over 4 M^2} \tilde F_3 \nonumber  \\
F'_2 &=& \tilde F_2  \nonumber \\
G'_A &=& -\tau \tilde F_3    \nonumber \\
G'_E &=& \tilde G_E +  {s-u \over 4 M^2} \tilde F_3     \ .
\ea

\noindent The invariants may be separated into parts coming from one-photon exchange and parts from two- or more-photon exchange,
\ba
G'_M &=& G_M + \delta  G'_M  \ ,    \nonumber \\
G'_E &=& G_E + \delta  G'_E  \ ,    \nonumber \\
G'_A &=& \delta G'_A  \ ,
\ea

\noindent where $G_M(Q^2)$ and $G_E(Q^2)$ are the usual magnetic and electric form factors, defined from matrix elements of the electromagnetic current and real for spacelike $Q^2$.  The quantities $\delta  G'_M$, $\delta  G'_E$, and $G'_A$ are ${\cal O}(e^2)$ relative to $G_M$ or $G_E$.

The reduced cross section in Eq.~(\ref{eq:reduced}) is
\begin{eqnarray}
\sigma_R 
= |G'_M|^2 + {\varepsilon \over \tau} |G'_E|^2 
		+ 2 \sqrt{ (1+\tau) (1-\varepsilon^2)  \over \tau } 
										\ G_M {\,\cal R\,}(G'_A)
		+  {\cal O}(e^4)  \ .
\end{eqnarray}

\noindent  The polarizations of the outgoing nucleons or analyzing powers of the target nucleons are
\ba
P_n &=& A_n =   \sqrt{\frac{2 \, \varepsilon \, (1+\varepsilon )}{\tau}} \,\,
	\frac{1}{\sigma_R} 
\left\{ {\cal I}(G^{\prime *}_E G'_M ) + 
		\sqrt{ { 1+\tau \over \tau }\cdot {1-\varepsilon \over 1+ \varepsilon} }
	\, G_E \, {\cal I} (G'_A) + {\cal O}(e^4)  \right\} \,,
										\nonumber	\\[1.75ex]
P_s &=& A_s = -\, P_e
	\sqrt{\frac{2\varepsilon (1 - \varepsilon)}{\tau}} \, \frac{1}{\sigma_R}
		\left\{ {\,\cal R\,} \left( G^{\prime *}_E G'_M  \right)
	+ \sqrt{ { 1+\tau \over \tau }\cdot {1+\varepsilon \over 1- \varepsilon} }
				\, G_E {\,\cal R\,} \left( G'_A \right) +  {\cal O}(e^4)
	\right\}   \,,
										\nonumber 	\\[1.75ex]
P_l &=& - A_l =  P_e
	 \,\frac{1}{\sigma_R}
	\left\{ \sqrt{1 - \varepsilon^2} \, |G'_M|^2 
		+ 2 \sqrt{ { 1+\tau \over \tau } } 
			\, G_M {\,\cal R}(G'_A) + {\cal O}(e^4)
	\right\}  \,.
\ea

\noindent The only single spin asymmetry is $P_n$ or $A_n$.  Further, $P_n$ or $A_n$ is zero if there be only one-photon exchange, so observation of a non-zero value is definitive evidence for multiple-photon exchange.  Polarizations $P_s$ or $P_l$ are double polarizations. The expressions for them are proportional to the electron longitudinal polarization $P_e$ (with, {\it e.g.}, $P_e =1$ if $h = +1/2$).


\subsection{Electron-quark elastic scattering amplitudes}


The two-photon part of electron-quark elastic scattering is given by
\ba
H_{h,\lambda} &=&
\frac{(e \, e_q)^2}{Q^{2}} \Big\{
	g_M \  \bar{u}(k', h)\gamma _{\mu }u(k, h) \,\cdot \, 
		\bar{u} (p'_q, \lambda) \gamma ^{\mu } u(p_q, \lambda)
										\nonumber \\
	&& \qquad + \ g_A^{(2\gamma)} \ 
		\bar{u}(k', h)\gamma _{\mu } \gamma_5 u(k, h) \,\cdot \, 
		\bar{u} (p'_q, \lambda) \gamma ^{\mu } \gamma^5 u(p_q, \lambda)
	\Big\} 
											\nonumber \\
	&=& \frac{(e \, e_q)^2}{Q^{2}} \Big\{
		\Big( \hat s - \hat u - (2h)(2\lambda) t \Big) \, g_M
		+ \Big( (2h)(2\lambda) (\hat s - \hat u) - t \Big) g_A^{(2\gamma)} 
		\Big\}  \ ,
\ea

\noindent when the electrons and quarks are both massless.  The theorem of Eq.~(\ref{eq:theorem}) relates
\ba
g_M &=& \tilde f_1 + { \hat s - \hat u \over 4 } \tilde f_3 \ ,  \nonumber \\
g_A^{(2\gamma)} &=& { t \over 4 } \tilde f_3  \ .
\ea

We split the two-photon part of $g_M$ into a hard and soft part, 
$g_M^{(2\gamma)} = g_M^{\,soft} + g_M^{\,hard}$, using the prescription of Grammer and Yennie~\cite{grammer}, and have
\ba
{\cal R} \left( g_M^{\,soft} \right) &=&  {\alpha \over \pi }
					\left\{ \ln \left( \frac{\lambda^2}{\sqrt{- \hat s \hat u}} \right) 
\ln \left( \frac{\hat s}{-\hat u} \right) + \frac{\pi^2}{2} \right\} \ ,
												\nonumber \\
{\cal I} \left( g_M^{\,soft} \right) &=& 
					\alpha \ln \left(\hat s \over \lambda^2 \right) \ ,
												\nonumber \\
{\cal R} \left( g_M^{\,hard} \right) &=&
	{ \alpha \over 4\pi } \Bigg\{ {-t \over \hat u} 
		\ln\left( \hat s \over -t \right)
			+ { t \over \hat s } \ln\left( \hat u \over t \right)
												\nonumber \\
	&& \qquad + \, { \hat s^2 + 3\hat u^2 \over 2\hat u^2 } 
		\ln^2\left( \hat s \over -t \right)
			- { 3\hat s^2 + \hat u^2 \over 2\hat s^2 } 
				\left( \ln^2\left( \hat u \over t \right) + \pi^2 \right)
					\Bigg\}  \ ,
												\nonumber \\
{\cal I} \left( g_M^{\,hard} \right) &=&
	-\alpha \left\{ { \hat s^2 + 3\hat u^2 \over 4\hat u^2 }
		\ln\left( \hat s \over -t \right) - { t \over 4\hat u } \right\}  \ ,
												\nonumber \\
{\cal R} \left( g_M^{(2\gamma)} \right) &=&  
	\frac{\alpha}{4\pi} \, \frac{t}{ \hat s \, \hat u} \,
		\left\{ \hat s \, \ln \left( \frac{\hat s}{-t}  \right) + 
			\hat u \, \ln \left( \frac{\hat u}{t}  \right) 
				+ \frac{\hat s - \hat u}{2} 
	\left[ \frac{\hat s}{\hat u} \ln^2 \left( \frac{\hat s}{-t}  \right)
		- \frac{\hat u}{\hat s} \ln^2 \left( \frac{\hat u}{t}  \right) 
			- \frac{\hat u}{\hat s} \pi^2
				\right] \right\} \ ,
												\nonumber \\
{\cal I} \left( g_M^{(2\gamma)} \right) &=&
	- \alpha
		\, \frac{t}{4 \hat u} \, \left\{ 
			\frac{\hat s - \hat u}{\hat u}  
				\ln \left( \frac{\hat s}{-t}  \right) + 1 \right\}  \ .
\ea


\subsection{Embedding}


The nucleon form factors are given in terms of the quark-level amplitudes and generalized parton distributions by
\ba											\label{eq:primedFF}
\delta G^{\prime(hard)}_M &=& { 1+\varepsilon \over 2\varepsilon } A 
			- { 1-\varepsilon \over 2\varepsilon } C   \ ,
											\nonumber \\
\delta G^{\prime(hard)}_E &=& \sqrt{ 1+\varepsilon \over 2\varepsilon }  B   \ ,
											\nonumber \\
\delta G^{\prime(hard)}_A &=& {t \over s-u }{ 1+\varepsilon \over 2\varepsilon }
			\left( A-C \right)  \ .
\ea

\noindent Quantities $A$, $B$, and $C$ are the same as in the text, but now written as
\ba
A &=& \int_{-1}^1 \frac{dx}{x}  \    
	{    (\hat s - \hat u) g_M^{\,hard} - t g_A^{(2\gamma)} 
			\over s-u    }
	\sum_q e_q^2 \, \left( H^q + E^q \right), 
											\nonumber \\
B &=& \int_{-1}^1 \frac{dx}{x}  \    
	{    (\hat s - \hat u) g_M^{\,hard} - t g_A^{(2\gamma)} 
			\over s-u    }
	\sum_q e_q^2 \, \left( H^q - \tau E^q \right), 
											\nonumber \\
C &=& \int_{-1}^1 \frac{dx}{x} \  
	{    (\hat s - \hat u) g_A^{(2\gamma)} - t g_M^{\,hard} 
			\over -t   }
	\, \mathrm{sgm}(x) \, 
	\sum_q e_q^2 \, \tilde H^q, 
\ea

\noindent and it is understood in Eq.~(\ref{eq:primedFF}) that the partonic amplitude $g_M$ has its soft part removed.



\section{Quark mass sensitivity}

\label{sec:withmass}



\subsection{Kinematics, and imaginary parts of the hard amplitudes}


We have until now set the quark mass to zero.

To investigate how severe this approximation is,  we will examine the effect of restoring the quark mass for the analyzing power calculations, though still keeping only the quark chirality conserving amplitudes.  There are three modifications.   The expressions for $\hat s$ and $\hat u$ become
\begin{eqnarray} 
\label{eq:withmass}
\hat s = \frac{(x + \eta)^2}{4 \, x \, \eta} \, Q^2 
		+ \frac{x+\eta}{x} m_q^2
\, , \qquad 
\hat u = - \frac{(x - \eta)^2}{4 \, x \, \eta} \, Q^2 
		+ \frac{x-\eta}{x} m_q^2   \,,
\end{eqnarray}
where $m_q$ is the effective quark mass.  The general electron-quark scattering amplitude, Eq.~(\ref{eq:tmatrixhard}), should have another term with a scalar function that we may call $\tilde f_2$ in analogy with the expansion of the electron-nucleon amplitude given in Eq.~(\ref{eq:tmatrix}).   However, this term flips quark helicities, and presently the formalism for embedding quark amplitudes into the nucleon using GPD's involves only the non-chirality flip GPD's.  There is neither theoretical development nor experimental information regarding chirality flip GPD's, and so we shall ignore $\tilde f_2$ as well as helicity flip parts of other amplitudes.  Including the quark mass leads to a modification of the hard scattering amplitudes so that
\ba
\frac{1}{2}
\left[ H_{h,\, + \frac{1}{2}}^{hard} + H_{h,\,- \frac{1}{2}}^{hard} \right]
	&\stackrel{(m_q = 0)}{=}& \frac{e^2}{Q^2} 
		\left\{ 
		\left[ \hat s - \hat u \right]
		\tilde f_1^{hard} 
		 - \hat s \hat u   \tilde f_3 
		\right\}						\nonumber \\
	&\stackrel{\to}{=}& \frac{e^2}{Q^2} 
		\left\{ 
		\left[ \hat s - \hat u - \frac{2m_q^2}{\hat s -m_q^2} Q^2 \right]
		\tilde f_1^{hard} 
		+ \left[m_q^4 - \hat s \hat u \right] \tilde f_3 
		\right\} \ ,
\ea
within the quantities $A$ and $B$ ($C$ is not needed for the analyzing power), and 
\ba
{\cal I}\left( \tilde{f}_1^{hard} \right)
&=& - \frac{e^2}{4 \pi} 
	\left\{ 
		\frac{ \frac{1}{2} \hat s  Q^2}
		{ \hat s Q^2  - \left( \hat s - m_q^2 \right)^2 }  \, 
		\ln \left[ \frac{\left( \hat s - m_q^2 \right)^2}{ \hat s Q^2}  \right] 
		+ \frac{ \hat s + m_q^2 }{2 \hat s} 
	\right\}, 
											\nonumber \\
{\cal I}\left( \tilde{f}_3 \right)
&=& - \frac{e^2}{4 \pi} \,
	\frac{\hat s - m_q^2 }{ \hat s Q^2  - \left( \hat s - m_q^2 \right)^2 } 
											\nonumber \\
		&&\times \ 
		\left\{ 
			\frac{ (\hat s - \hat u) ( \hat s - m_q^2) }
				{ \hat s Q^2  - \left( \hat s - m_q^2 \right)^2 }  \,
		\ln \left[ \frac{\left( \hat s - m_q^2 \right)^2}{ \hat s Q^2}  \right] 
			+ \frac{ \hat s + m_q^2 }{\hat s} 
		\right\} \,.
\ea


\begin{figure}
\includegraphics[width=7.8cm]{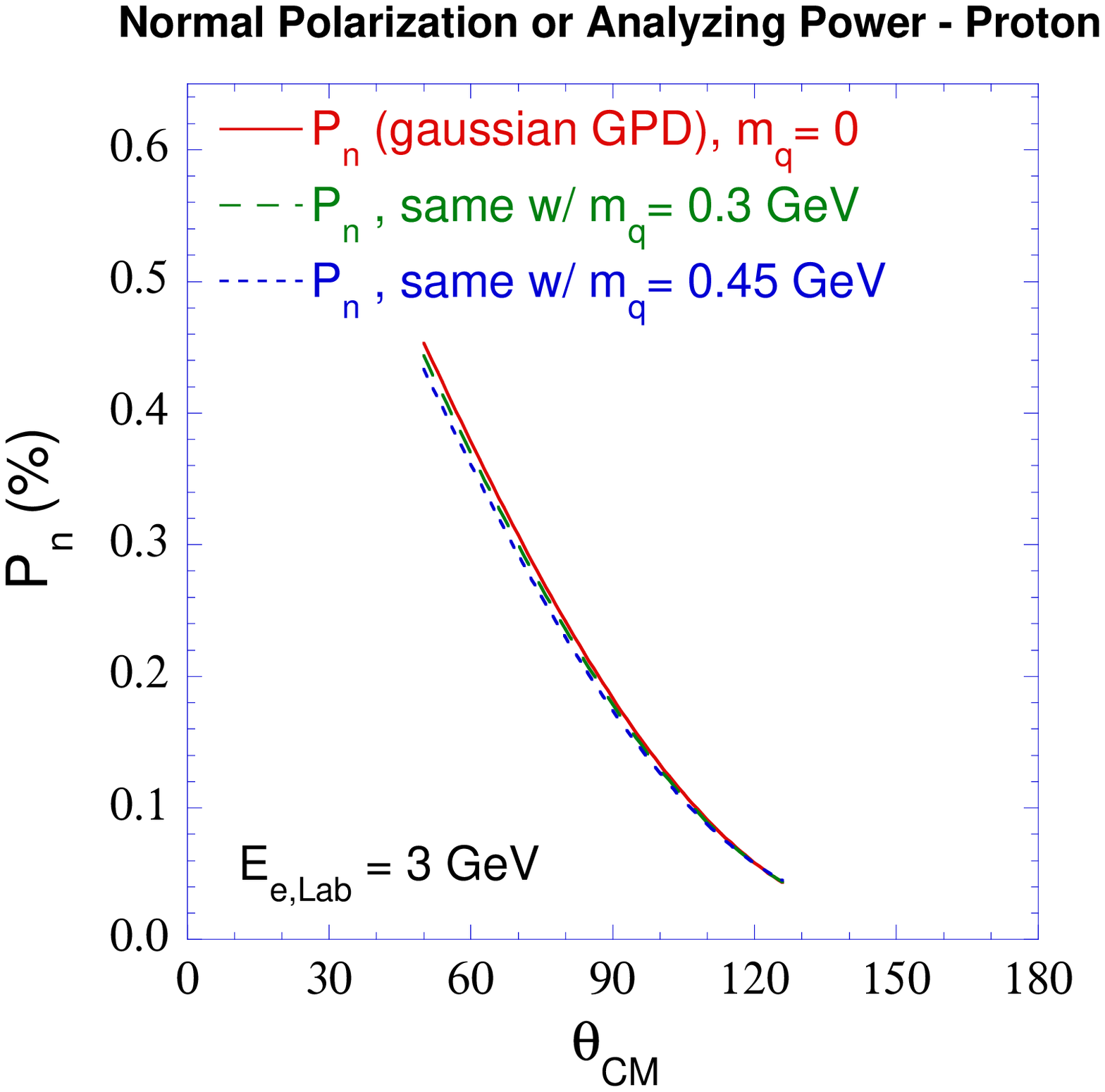}
\hfill
\includegraphics[width=7.8cm]{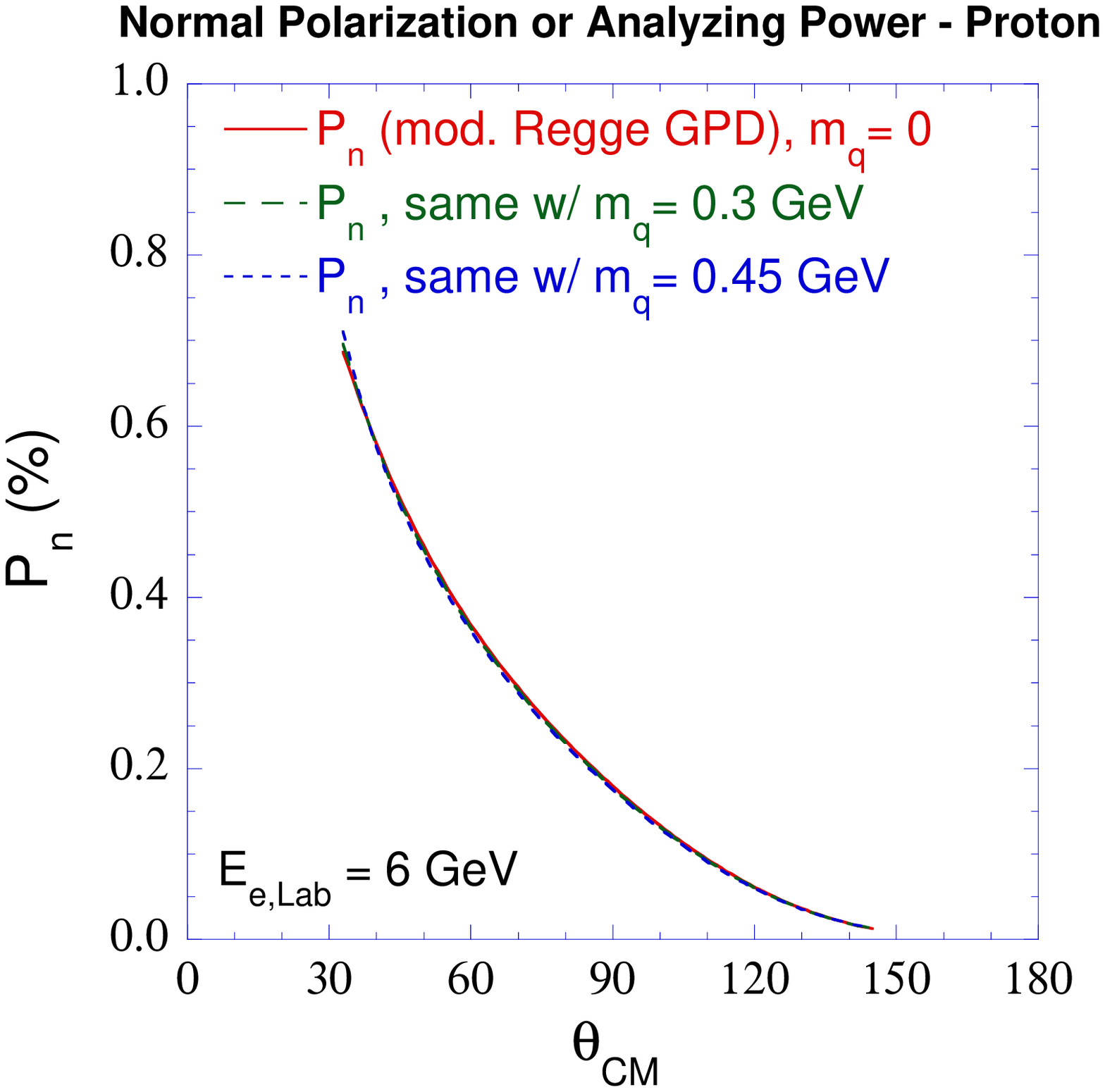}

\caption{Quark mass correction plots for the analyzing power or normal polarization, for fixed electron incoming lab energies of 3 and 6 GeV.
}
\label{fig:quark_masses}
\end{figure}


The results for the analyzing power $A_n$ when including quark masses in the quark helicity conserving amplitudes are shown in Fig.~\ref{fig:quark_masses} for quark masses 300 MeV and 450 MeV.  The effects are clearly not large.


\subsection{Real parts of hard two-photon exchange amplitudes with finite quark mass}


When the quark mass is not zero, we have

\begin{eqnarray}
{\cal R} \left(\tilde f_1^{hard} \right)
&=& \frac{e^2}{4 \pi^2} \cdot  \left\{
\left[
\ln\left(\frac{\sqrt{(\hat s - m_q^2)|\hat u - m_q^2|}}{Q^2}\right)
\,+\, \frac{1}{2} \right]
\cdot \ln\left| \frac{\hat{s}-m_q^2}{\hat u - m_q^2}\right|
\right. \nonumber \\
&&+\frac{m_q^2}{2}\left[
\frac{1}{\hat{s}}\ln\left(\frac{\hat{s}-m_q^2}{m_q^2}\right)
-\frac{1}{\hat{u}}\ln\left|\frac{\hat{u} - m_q^2}{m_q^2}\right|\right]
\nonumber \\
&&+\frac{1}{2} \frac{Q^2 (\hat{s}-\hat{u})}{(m_q^4-\hat{s}\hat{u})}
\left[ \, \frac{\pi^2}{2}
	+\frac{1}{4} \ln^2\left(\frac{m_q^2}{Q^2}\right) \right]
		-\frac{1}{2} \ln^2\left(\frac{\hat{s}-m_q^2}{Q^2}\right)
			+\frac{1}{2} \ln^2\left|\frac{\hat{u}-m_q^2}{Q^2}\right| 
\nonumber \\
&&
+ \frac{1}{2}
\frac{(\hat s - \hat u)(\hat{s}+\hat{u})}{(m_q^4 - \hat s \hat u)}
\, \frac{1}{\chi}
					\nonumber \\
&& \times
\left[\,
	L\left(\frac{2}{1+\chi}\right) - \frac{\pi^2}{2}
		+ \frac{1}{2}
\ln\left(\frac{m_q^2}{Q^2}\right)\ln\left(\frac{1+\chi}{-1+\chi}\right)
		+ \frac{1}{4}\ln^2\left(\frac{1+\chi}{-1 + \chi}\right)
				\right]  
					\nonumber \\
&&
- \frac{1}{2} \frac{Q^2 \, \hat{s}}{(m_q^4-\hat{s}\hat{u})}
\left[L\left(\frac{\hat s - m_q^2}{\hat{s}}\right)
 + \frac{1}{2}\ln^2\left(\frac{\hat{s}-m_q^2}{\hat{s}}\right)
 + \frac{1}{2}\ln^2\left(\frac{\hat{s}-m_q^2}{Q^2}\right)
\right] \nonumber \\ 
&& + \frac{1}{2} \frac{Q^2 \,\hat{u}}{(m_q^4-\hat{s}\hat{u})} 
\left[ 
	- L\left(\frac{\hat u }{\hat{u} - m_q^2 } \right) 
		 +\frac{5\pi^2}{6}  
		 	+ \frac{1}{2}\ln^2\left|\frac{\hat u - m_q^2}{Q^2}\right|
\right]
\Bigg\}    \,,       
\end{eqnarray}

\noindent and

\begin{eqnarray}
{\cal R} \left( \tilde f_3 \right)
&=& - \frac{e^2}{4 \pi^2} \frac{1}{(m_q^4-\hat{s}\hat{u})} \cdot
	\left\{
	\frac{\hat{s}^2-m_q^4}{\hat{s}}\ln\left(\frac{\hat{s}-m_q^2}{Q^2}\right)
	+\frac{\hat{u}^2-m_q^4}{\hat{u}}\ln\left|\frac{\hat u - m_q^2}{Q^2}\right|
\right.
\nonumber \\
&&+ \frac{(\hat{s}-\hat{u})^2 Q^2}{(m_q^4-\hat{s}\hat{u})}
	\left[ \frac{\pi^2}{3} +\frac{1}{4}\ln^2\left(\frac{m_q^2}{Q^2}\right)\right]
+ m_q^2 \left( -2 + \frac{m_q^2 (\hat{s}+\hat{u})}{\hat{s}\hat{u}} \right)
\ln\left(\frac{m_q^2}{Q^2}\right)
\nonumber \\
&&+
\frac{(\hat{s}-\hat{u})^2 (\hat{s}+\hat{u}) + 4m_q^2 (m_q^4-\hat{s}\hat{u})}
{  (m_q^4-\hat{s}\hat{u})}  \,  \frac{1}{\chi} 
\nonumber \\
&&\times \left[ \,
	L\left(\frac{2}{1+\chi}\right) - \frac{\pi^2}{2}
		+ \frac{1}{2}
\ln\left(\frac{m_q^2}{Q^2}\right)\ln\left(\frac{1+\chi}{-1+\chi}\right)
		+ \frac{1}{4}\ln^2\left(\frac{1+\chi}{-1 + \chi}\right)
			\right]
\nonumber \\
&&
+\frac{(\hat{s}-\hat{u})(\hat{s}-m_q^2)^2}{(m_q^4-\hat{s}\hat{u})}
\left[- L \left(\frac{\hat s - m_q^2}{\hat{s}}\right)
 +\frac{\pi^2}{6} - \frac{1}{2}\ln^2\left(\frac{\hat{s}-m_q^2}{\hat{s}}\right) -\frac{1}{2}\ln^2\left(\frac{\hat{s}-m_q^2}{Q^2}\right)\right]
\nonumber \\
&&
+\frac{(\hat{s}-\hat{u})(\hat{u} - m_q^2)^2}{(m_q^4-\hat{s}\hat{u})}
\left.   \left[ 
	- L\left(\frac{\hat u }{\hat{u} - m_q^2 } \right) 
		 +\frac{2\pi^2}{3}  
		 	+ \frac{1}{2}\ln^2\left|\frac{\hat u - m_q^2}{Q^2}\right|
\right] \right\},
\end{eqnarray}

\noindent where $\hat s$ and $\hat u$ were given in  Eq.~(\ref{eq:withmass}), and
\begin{eqnarray}
	\chi \equiv \sqrt{1 + 4 m_q^2 / Q^2} \,.
\end{eqnarray}


\begin{figure}
\includegraphics[width=7.8cm]{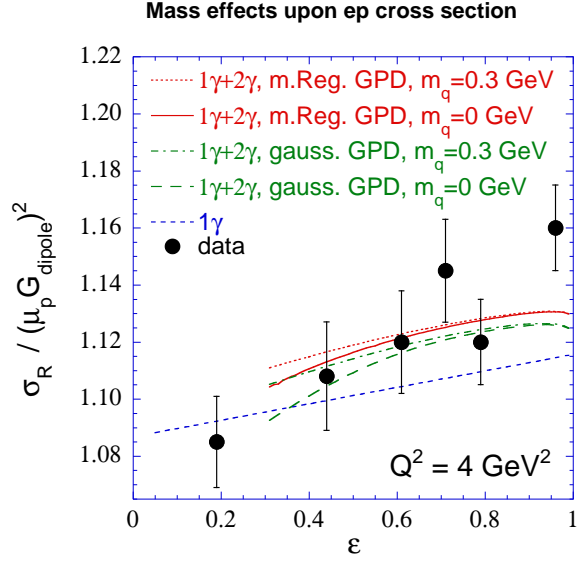}

\caption{Quark mass correction plot for the reduced cross section.  The curves are as labeled.  In this plot, all curves have $G_M$ set to 0.99 times the Brash {\it et al.}\ value~\cite{Bra02}.  The data is from~\cite{Slac94}.
}
\label{fig:quark_masses_real}
\end{figure}


The effect of the quark mass corrections upon the reduced cross section is shown in Fig.~\ref{fig:quark_masses_real} for $Q^2 = 4$ GeV$^2$.  One sees from Fig.~\ref{fig:quark_masses_real} that the quark mass effects mainly influence our result at small values of $\varepsilon$, where $|u|$ becomes small.  They show the theoretical error on our calculation in this region. A full calculation also requires quantifying the effect of the cat's ears diagrams.  
  A study of such corrections is clearly worthwhile for a future work, both for two-photon exchange amplitudes and for wide-angle Compton scattering.

\end{document}